\documentclass[11pt]{article}
\usepackage[toc,page]{appendix}
\usepackage{graphicx,color} 
\graphicspath{{figures/}}
\usepackage{jheppub}
\usepackage{amsmath}
\usepackage{amssymb}
\usepackage{subcaption}
\usepackage{enumitem}
\usepackage{mathtools}
\usepackage{comment}
\usepackage[utf8]{inputenc}
\usepackage{tikz}
\usetikzlibrary{decorations.pathreplacing}
\usepackage{parskip}
\definecolor{rust}{rgb}{0.8,0.2,0.2}
\definecolor{alizarin}{rgb}{0.82, 0.1, 0.26}
\definecolor{palecarmine}{rgb}{0.69, 0.25, 0.21}
\definecolor{lightblue}{rgb}{.1,.4,.5}
\definecolor{brown}{rgb}{.64,.43,.138}

\def \rm{r_{min}}
\DeclareMathOperator{\csch}{csch}
\def \Lm{L_{-1}}

\newcommand{\bea}{\begin{eqnarray}}
\newcommand{\eea}{\end{eqnarray}}
\newcommand{\be}{\begin{equation}}
\newcommand{\ee}{\end{equation}}
\newcommand{\ba}{\begin{align}}
\newcommand{\ea}{\end{align}}

\newcommand{\nn}{\nonumber\\}

\def\br{{\beta_R}}

\def\l1{{{1-loop}}}

\def\bz{{\bar{z}}}
\def\by{{\bar{y}}}

\def\n1{\Bigg|_{n=1}}

\def\n{{(n)}}

\newcommand\sig{\sigma}

\newcommand\om{\omega}

\def\le{\left}
\newcommand\ri{\right}

\newcommand{\bbgv}{Bogoliubov }

\newcommand{\ha}{\frac{1}{2}}

\title{Imprints of dynamical phases in semiclassical entanglement entropy in 2D CFT }

\author[a]{Parijat Dey} \emailAdd{parijat.dey@bose.res.in}
\author[a]{\!\!, Semanti Dutta} \emailAdd{semanti.dutta@bose.res.in}
\author[b]{and Bobby Ezhuthachan} \emailAdd{bobby.phy@gm.rkmvu.ac.in}
\affiliation[a]{Department of Astrophysics and High Energy Physics,\\
S.N. Bose National Centre for Basic Sciences,
Salt Lake, Kolkata 700106, India}
\affiliation[b]{Ramakrishna Mission Vivekananda Educational and Research Institute, Belur Math, Howrah 711202, West Bengal, India}	

\abstract{We study the time evolution of semiclassical 
entanglement entropy in a class of $sl(2,\mathbb{R})$ driven states in a large $c$ conformal field theory (CFT) 
in $1+1$ spacetime dimensions. 
Upon varying the parameters of the drive, we find that the entanglement entropy exhibits the signature of the dynamical phases of the driven CFT. We further study the holographic dual of this CFT where the excited states of a minimally coupled scalar in $AdS_3$ induce a backreaction that modifies the background geometry.
We compute the back reacted geometry by solving Einstein's equation with the expectation value of the stress tensor in the coherent state as the source term. Subsequently, we calculate the time evolution of the perturbed minimal area and the bulk entanglement entropy at $O(G_N^0)$, up to sub-leading order in the short distance approximation. These results match with the CFT entanglement entropy in the boundary at $O(c^0)$, and serve as a nontrivial check of the Faulkner-Lewkowycz-Maldacena (FLM) conjecture for the first quantum correction of holographic entanglement entropy. The details of the check has been provided in the accompanying ancillary notebook.}

\begin{document}
\maketitle

\section{Introduction}

One of the most interesting developments in the field of quantum gravity over the past two decades has been the discovery of a deep connection between entanglement in conformal field theories and geometry in a higher dimensional gravity theory \cite{Ryu:2006bv, Hubeny:2007xt}. Investigations into the nature of these connections have led to a deeper understanding of the fundamentals of holography, and have provided new insights into  unresolved problems, such as the black hole information paradox \cite{Mathur:2009hf} and the nature of the black hole interior \cite{Maldacena:2013xja}. Central to many of these developments has been the Quantum Extremal Surface (QES) formula \cite{Engelhardt:2014gca}, which relates the entanglement entropy (EE) of a subregion ($A$) in a CFT to the  ${\it generalized\; entropy}$ in the bulk.
\begin{equation}\label{QES}
   S_{bndy}(A) = \underset{\text{ext}}{\min}\Big[ \frac{Area(\gamma_{A})}{4G_N} + S_{bulk}(\Sigma_A)\Big]\,.
\end{equation}
The QES formula states that the entanglement entropy associated with the reduced density matrix in region $A$, can be obtained from the bulk description by taking the minimum value among all the extrema of the ${\it generalized\; entropy}$ \footnote {The term in the RHS inside the $square$ $bracket$ is denoted as ${\it generalized\; entropy}$ , which is the sum of the area of $\gamma_A$ and the entanglement entropy of the bulk fields in the region $\Sigma_A$, which is the bulk volume bounded by $\gamma_A$ and $A$. }  ($S_{gen}$), when varied over  $\gamma_A$, which  is a surface in the bulk, homologous to $A$. The  $\gamma_A$ for which $S_{gen}$ is minimized is called the Quantum Extremal surface. The domain of dependence of this region forms the entanglement wedge, i.e., the maximal spacetime region that can be re-constructable from the boundary subregion $A$ \cite{Dong:2016eik}. One significant application of equation (\ref{QES}) has been towards finding a resolution of the black hole information paradox, where incorporating the contribution from Quantum Extremal $Islands$ to the entanglement entropy of radiation led to the page curve directly from a gravity calculation \cite{Penington:2019npb, Almheiri:2019hni}.

At lowest order in $G_N$, the QES formula reduces to the Faulkner-Lewkowycz-Maldacena(FLM) correction to the Ryu-Takayangi (RT) formula for the holographic entanglement entropy.
\begin{equation}\label{FLM}
S_{bndy}(A) =  \underset{\text{ext}}{\min}\Big[\frac{Area(\gamma_A)}{4G_N}\Big] + S_{bulk}(\Sigma_A) +\mathcal{O}(G_N)\,,
\end{equation}
where $\gamma_A$ is the extremal surface over which the area $Area(\gamma_A)$ is minimum and $\Sigma_A$  is the region bounded by $\gamma_A$ and $A$.  Equation (\ref{QES}) and equation (\ref{FLM}) provide a quantifiable connection between quantization of the bulk  and boundary Hilbert spaces. For this reason, it would be nice to have explicit examples where one could check these formulae, as it provides a window into bulk quantization. Few such checks are in the literature due to the complexity of the computation \cite{Sugishita:2016iel,Belin:2018juv,Agon:2020fqs,Chowdhury:2024fpd,Bhat:2025iqb,Colin-Ellerin:2024npf,Colin-Ellerin:2025dgq}. In \cite{Belin:2018juv}, the authors have directly checked the conjecture for vacuum subtracted entanglement entropy of the primary excitation in minimally coupled scalar field in $AdS_3$. The same formalism was extended for the global descendants and simple superposed states in \cite{Chowdhury:2024fpd,Bhat:2025iqb}, and for vectors in the free Chern-Simons theory. In \cite{Chowdhury:2024fpd}, it was also noted that in the short distance approximation, the entanglement entropy of the descendants scales by a kinematic factor w.r.t the primary excitation manifesting the Virasoro symmetry. The same behavior of the entanglement entropy was observed earlier in CFT \cite{Chowdhury:2021qja}. In \cite{Agon:2020fqs}, the check was performed in a time-dependent setting  , where the authors considered an infalling scalar field in a  pure $AdS_3$ geometry, corresponding to a regularized quench in the CFT.

In this paper, we extend the list of examples to include two dimensional periodically driven CFTs \cite{Wen:2018agb,Wen:2020wee}, which have recently garnered traction as analytically tractable models to study novel phenomenon arising in non-equilibrium situations, both in the field theory, as well as through their bulk realizations via the holographic correspondence \cite{das:2022brane,Jiang:2024hgt}. In these models, the effect of an external drive is to deform the CFT Hamiltonian, by a conformal generator. Thus, the time evolution under such deformed Hamiltonians is equivalent to a  time dependent conformal map, making it possible to use the full power of the underlying conformal invariance. In the special case, when the Hamiltonian is deformed by the generators of the $sl(2,\mathbb{R})$ sub-algebra of the conformal algebra, the time evolution becomes particularly tractable. The stroboscopic dynamics of such $sl(2,\mathbb{R})$ driven CFTs induces three qualitatively distinct ${\it dynamical}$ ${\it phases}\footnote{These have been referred to in the literature as the heating phase, the non heating phase, and the phase boundary.}$, characterized by the qualitatively distinct temporal growth of entanglement entropy and total energy, in each phase\cite{Wen:2020wee}.

In this paper, following the methods introduced in \cite{Agon:2020fqs}, we study the time evolution of semiclassical entanglement entropy of an interval of size $2\pi x (x \ll 1)$ in 2D CFT on a ring with circumference $2\pi$, for a class of states:
\begin{equation}\label{drive state}
|\Psi(t)\rangle = e^{-it H_{0}}e^{-inTH_{eff}}|h,h'\rangle,
\end{equation}
where $H_{eff}$ is a $sl(2,\mathbb{R})$ valued Hamiltonian of the form: 
\begin{equation}
H_{eff} = \alpha L_{0} +\alpha'\bar{L}_0 +\beta (L_{-1} + L_{1})+\beta'(\bar{L}_{-1} +\bar{L}_1),
\end{equation}
and $nT$ denotes the stroboscopic time after $n$ iterations \cite{das:2022brane}. 
We compute the leading order ($O(c^{0})$) correction to the entanglement entropy for these states in the small interval limit for each dynamical phase of the deformed Hamiltonian which are classified by the sign of the discriminant $\alpha^2 -4\beta^2 $.

In the bulk $AdS_3$, the dual state of \eqref{drive state} is a superposition of the primary state and its descendants, which is constructed by applying isometry generators on the primary state. Hence we can extend the formalism developed in \cite{Chowdhury:2024fpd} to find the time evolution of the holographic entanglement entropy in order $G_N^0$ corresponding to the above mentioned subsystem in CFT \footnote{While in the bulk the time evolution occurs in time $t$, we can obtain the entanglement entropy correction to the driven CFT state ($e^{-inTH_{eff}}|h,h'\rangle$) at stroboscopic time $nT$, by simply taking $t=0$.}. The entanglement entropy matches that obtained in CFT, confirming the FLM conjecture. In both CFT and bulk dual, we have not considered the entanglement entropy at $O(c^1)$ or $O\left(\frac{1}{G_N}\right)$.

In CFT, our analysis remains valid for the general state given in equation (\ref{drive state}), the explicit results are presented for two specific cases (i) $(\alpha =\alpha',\; \beta =\beta'\; \textrm{and}\; h=h') $ and (ii) $(\alpha'=\beta'=0,\; \textrm{and}\; h=h')$. The details of these computations are presented in section \ref{2} and appendix \ref{appendix:mobius}, while some useful identities are collected in appendix \ref{bch}. In appendix \ref{appendix:drive}, for completeness, we also present a brief review of relevant results of driven CFTs.  The explicit expressions for the leading correction to the entanglement entropy in each phase of the drive are given in section \ref{sec:ee_3phases}. We see a qualitatively distinct dependence on the stroboscopic time $nT$. In the non-heating phase, one sees an oscillatory dependence on $nT$. In the heating phase, we obtain rapidly growing entropy for intervals containing the energy peak.

 In section \ref{3}, we present the computation in the bulk. The bulk computation is performed for the case of a holomorphic drive with $(\alpha =0,\; \alpha'=\beta'= 0, \; \textrm {and}\;h=h')$. For this case, we are able to check equation (\ref{FLM}) in the short distance approximation, following the methods developed in \cite{Belin:2018juv,Chowdhury:2024fpd}. In the bulk, the role of the drive is to prepare an initial state ($e^{-inTH_{eff}}|h,h\rangle$). This state is then evolved with the unperturbed CFT Hamiltonian. At $t=0$ we observed the signature of the Heating phase, where the entanglement entropy increases rapidly when varying with $n T \beta$ . However, this computation in bulk is valid only for small stroboscopic times $n T$ and the drive parameter $\beta$ in the heating phase and the phase boundary, as the energy density increases sharply with $nT \beta$. This is because we have assumed that the back reaction of the prepared state and the change in the reduced density matrix of the entanglement wedge w.r.t. the vacuum counterpart are small.

We present the key expressions of the computation of the entanglement entropy in the bulk as well as some checks of the same in several appendices  from  [\ref{appendix: D} - \ref{appendix I}]. We also provide an ancillary Mathematica notebook, containing the details of the computations in the bulk. Finally, we end with a discussion of our results and comments on future work in section \ref{discussion}.


\section{Entanglement entropy of excited states in CFT }\label{2}
\begin{figure}[!h]
\centering
\begin{tikzpicture}[scale=1.8]
\draw (-1.2,1.7)
arc[start angle=180,end angle=360,x radius=1.2,y radius=0.22];

\draw
(1.2,1.7)
arc[start angle=0,end angle=180,x radius=1.2,y radius=0.22];

\draw (-1.2,-1.7)
arc[start angle=180,end angle=360,x radius=1.2,y radius=0.22];

\draw
(1.2,-1.7)
arc[start angle=0,end angle=180,x radius=1.2,y radius=0.22];

\draw (-1.2,-1.7) -- (-1.2,1.7);
\draw (1.2,-1.7) -- (1.2,1.7);

\draw (-1.2,0)
arc[start angle=180,end angle=360,x radius=1.2,y radius=0.22];

\draw
(1.2,0)
arc[start angle=0,end angle=180,x radius=1.2,y radius=0.22];



\node[right] at (1.55,1.7) {$\tau=\infty$};

\node[right] at (1.55,0) {$\tau=0$};

\node[right] at (1.55,-1.7) {$\tau=-\infty$};


\node at (-0.15,-0.08) {$0$};

\node at (0.69,-0.08) {$2\pi x$};

\draw[fill=blue]
plot[smooth cycle] coordinates {
(0.02,{ -0.22*sqrt(1-(0.02/1.2)^2) + 0.03})
(0.06,{ -0.22*sqrt(1-(0.06/1.2)^2) + 0.06})
(0.10,{ -0.22*sqrt(1-(0.10/1.2)^2) + 0.02})
(0.14,{ -0.22*sqrt(1-(0.14/1.2)^2) + 0.06})
(0.18,{ -0.22*sqrt(1-(0.18/1.2)^2) + 0.02})
(0.22,{ -0.22*sqrt(1-(0.22/1.2)^2) + 0.06})
(0.26,{ -0.22*sqrt(1-(0.26/1.2)^2) + 0.02})
(0.30,{ -0.22*sqrt(1-(0.30/1.2)^2) + 0.06})
(0.34,{ -0.22*sqrt(1-(0.34/1.2)^2) + 0.02})
(0.38,{ -0.22*sqrt(1-(0.38/1.2)^2) + 0.06})
(0.42,{ -0.22*sqrt(1-(0.42/1.2)^2) + 0.02})
(0.46,{ -0.22*sqrt(1-(0.46/1.2)^2) + 0.06})
(0.46,{ -0.22*sqrt(1-(0.46/1.2)^2) - 0.02})
(0.42,{ -0.22*sqrt(1-(0.42/1.2)^2) - 0.06})
(0.38,{ -0.22*sqrt(1-(0.38/1.2)^2) - 0.02})
(0.34,{ -0.22*sqrt(1-(0.34/1.2)^2) - 0.06})
(0.30,{ -0.22*sqrt(1-(0.30/1.2)^2) - 0.02})
(0.26,{ -0.22*sqrt(1-(0.26/1.2)^2) - 0.06})
(0.22,{ -0.22*sqrt(1-(0.22/1.2)^2) - 0.02})
(0.18,{ -0.22*sqrt(1-(0.18/1.2)^2) - 0.06})
(0.14,{ -0.22*sqrt(1-(0.14/1.2)^2) - 0.02})
(0.10,{ -0.22*sqrt(1-(0.10/1.2)^2) - 0.06})
(0.06,{ -0.22*sqrt(1-(0.06/1.2)^2) - 0.02})
};

\draw[->,thick]
(1.95,-1) -- (1.95,-0.5)
node[midway,right] {$\tau$};

\draw[<-,thick]
(0.75,-0.33)
arc[start angle=-10,end angle=-170,
    x radius=0.75,
    y radius=0.14];

\node at (0,-0.55) {$\varphi$};
\end{tikzpicture}
\caption{A CFT living on a cylinder labeled by the coordinates $(\tau,\varphi)$. The entangling region $A$, an interval  of length  $ [0, 2\pi x]$ on the spatial circle of the cylinder at $\tau=0$  is denoted by the blue cut.}\label{fig:cylinder}
\end{figure}
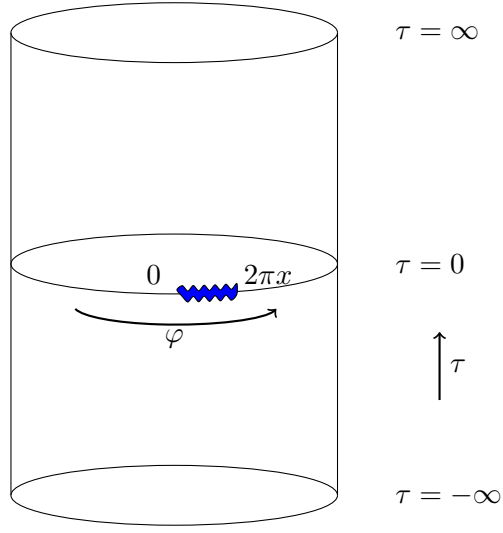

In this section, we study the time evolution of the entanglement entropy in an excited  state using replica trick \cite{Calabrese:2004eu, Calabrese:2009qy, Alcaraz:2011tn, Sarosi:2016oks}, in a 2D CFT living on a cylinder. {\it We first map the cylinder to the Euclidean plane, after doing a Wick rotation.  At the end of the computation, we analytically continue back to real time}. In Euclidean signature, the cylinder $\mathbb{R}\times S^1$ is labelled by the coordinates $(\tau,\varphi)  \sim (\tau,\varphi+2\pi)$.  We want to compute the time evolution of  entanglement entropy between a region $A$ and its complement. We choose the entangling region $A$ to be an interval  of length  $ [0, 2\pi x]$ on the spatial circle of the cylinder at $\tau=0$ as shown in Figure \ref{fig:cylinder}. 

It is convenient to map the cylinder to the complex plane $\mathbb{C}$ with coordinates $(z,\bar{z})$ using the following transformation
\begin{align}
    z&=e^{\tau+i \varphi}\,,\quad \bar{z} =e^{\tau-i \varphi}\,.
\end{align}
We are interested in the following CFT state 
\begin{align}\label{eqn:state}
    |\psi(\tau)\rangle & = e^{- H_0 \tau} |\Psi_{boundary}(n T)\rangle\,,
\end{align}
where $H_0 =L_0+\bar{L}_0$. The state $|\Psi_{boundary}(nT)\rangle$ is prepared by evolving a  primary state of weight $(h,h)$, under a periodic two step discrete drive (see appendix \ref{appendix:drive} for a review of a periodically driven CFT), with periodicity $T$ for $n$ cycles, i.e., for a stroboscopic time $n T$ (for brevity of notation, we denote the Euclidean stroboscopic time also as $T$, at the end we will take $T \rightarrow i T$)
\begin{align}\label{eqn:cft_bdystate}
    |\Psi_{boundary}(n T)\rangle=e^{-n T H_{eff}}|h,h\rangle\,,
\end{align}
where
\begin{align}\label{eq: H eff CFT main}
    H_{eff}&=\alpha (L_0+\bar{L}_0)+\beta(L_{-1}+L_1+\bar{L}_{-1}+\bar{L}_1)\,.
\end{align}
Here $\alpha, \beta$ are real parameters that ensure the Hermiticity of $H_{eff}$ in Lorentzian time\footnote{Without loss of generality, in our examples, we choose $(\alpha,\; \beta) \geq 0$.}. Depending on these parameters, one can have the following three dynamical phases of the system
\begin{equation}\label{eqn:3phases}
\tag{\theequation}
\begin{minipage}{0.9\linewidth}
\begin{enumerate}[label={(\alph*)}]
\item heating phase ($\alpha^2 < 4\beta^2$),
\item non-heating phase ($\alpha^2 > 4\beta^2$) and
\item phase boundary ($\alpha^2 = 4\beta^2$).
\end{enumerate}
\end{minipage}
\end{equation}
For later purposes, we find it useful to define $\delta^2$, the $sl(2,\mathbb{R})$ invariant as follows:
\begin{align}\label{eqn:delta_def}
\delta^2 &\equiv 4\beta^2-\alpha^2 \,: \quad \text{heating phase} \,,\nn
    \delta^2 &\equiv \alpha^2 -4\beta^2\,: \quad \text{non heating phase} \,,\nn
\alpha & = 2\beta\,: \quad \text{phase boundary} \,.
\end{align}
We suppress the label $nT$ in \eqref{eqn:cft_bdystate} and denote the state by $|\Psi_{boundary}\rangle$ in what follows. Once we prepare this state, we stop the drive and let the state  evolve under the CFT Hamiltonian $H_0$ for all $\tau \geq n T$. The state $|h,h\rangle$ can be obtained by acting with a Virasoro primary operator $\mathcal{O}$ of dimension $(h,h)$ on the vacuum as
\begin{align}
    |h,h\rangle=\mathcal{O}(0,0)|0\rangle\,.
\end{align}
The conjugate state is given by
\begin{align}
  \langle h,h |  =\underset{z,\bar{z}\rightarrow \infty}{\lim}\langle 0|\mathcal{O}(z,\bar{z})(z \bar{z})^{2h}\,.
\end{align}
One can express $|\psi(\tau)\rangle$  in \eqref{eqn:state} as follows 
\begin{equation}\label{eqn:psitau}
 |\psi(\tau)\rangle = \mathcal{N} \mathcal{O}(z_1,\bar{z}_1)|0\rangle \, ,
\end{equation}
where 
the location of the operator $(z_1, \bar{z}_1)$, is obtained by applying the successive conformal transformations generated by $e^{-\tau H_0}e^{-nT H_{eff}}$, while the transformation factors have been absorbed into the normalization $\mathcal{N}$ (see Appendix \ref{appendix:mobius} for details).   

\subsection{Replica method}
In this subsection, we compute the entanglement entropy of the state \eqref{eqn:psitau} following the method discussed in  \cite{Agon:2020fqs}. 
In order to compute the entanglement entropy, we define the reduced density matrix at time $\tau$ as
\begin{align}
    \rho_A(\tau) &\equiv \text{Tr}_{\bar{A}} |\psi(\tau)\rangle \langle\psi(\tau)|\,\nn
   & = \mathcal{N}\mathcal{N^*}\mathcal{O}(z_1,\bar{z}_1)|0\rangle\langle 0| \mathcal{O}^{\dagger}(z_1, \bar{z}_1)
    \footnotemark \,,
\end{align}
\footnotetext{Here, $\mathcal{O}^{\dagger}(z_1,\bar{z}_1) \equiv (z_1\bar{z}_1)^{-2h}\mathcal{O}(\frac{1}{\bar{z}_1},\frac{1}{z_1})$. The normalization factors will cancel in the final computation of the EE, and so we do not write down the explicit form of these here.} 
where $\bar{A}$ is the complement of the region $A$.
We would like to calculate  the Von Neumann entropy of the reduced density matrix
\begin{align}
    S_{EE}(\tau)=-\text{Tr}{\rho_{A}}(\tau)\log \rho_{A}(\tau)\,.
\end{align}
It is useful to first compute the Renyi entropy $S_q$ and then compute $S_{EE}(\tau)$ using:
\begin{align}
    S_{EE}(\tau)= \lim_{q\rightarrow1} S_q\,,
\end{align}
where
\begin{align}
    S_q \equiv \frac{1}{1-q} \log\text{Tr}\rho^q_A(\tau)\,.
\end{align}
Since the entanglement entropy is UV divergent, we focus on the regularised entanglement entropy which is obtained by subtracting the contribution from the vacuum without any operator insertion, which is as follows
\begin{align}
      \delta S_{EE}(\tau)\equiv \lim_{q\rightarrow1} S_q-S^{vac}_q\,.
\end{align}
The subsystem $A$ on the $q$ sheeted cylinder is mapped to the $q$ complex sheeted plane $\Sigma_q$ whose endpoints are
\begin{align}\label{eqn:interval}
    u=e^{2\pi i x}, \quad v=1\,.
\end{align}
The entanglement entropy can be written in terms of a $2q$-point correlation function on $\Sigma_q$.
\begin{align}\label{eqn:ee2}
   \delta S_{EE}(\tau)= \lim_{q\rightarrow1}  \frac{1}{1-q}\log \bigg[\frac{\langle \prod_{k=0}^{q-1}\mathcal{O}(z^{(k)}_1, \bar{z}^{(k)}_1) \mathcal{O}^\dagger(z^{(k)}_1, \bar{z}^{(k)}_1)\rangle_{{{\Sigma}}_q}}{{\langle \mathcal{O}(z^{(0)}_1, \bar{z}^{(0)}_1) \mathcal{O}^\dagger(z^{(0)}_1, \bar{z}^{(0)}_1)\rangle}^q _\mathbb{C}}\bigg]\,.
\end{align}
The operator locations are denoted by $(z^{(k)}_1, \bar{z}^{(k)}_1)$  on the $k$-th sheet. 
In general, it is difficult to compute the $2q$-point correlation function on $\Sigma_q$. This can be simplified by making use of the conformal transformations in two dimensions and using the {\it uniformization map} from $\Sigma_q \rightarrow \mathbb{C}$
\begin{align}
    w=\bigg(\frac{z-u}{z-v}\bigg)^{\frac{1}{q}}\,,\quad \bar{w}=\bigg(\frac{\bar{z}-\bar{u}}{\bar{z}-v}\bigg)^{\frac{1}{q}}\,,
\end{align}
where $z\in \Sigma_q$ and $w\in \mathbb{C}$. This transformation maps the left and right ends of the entangling region to the origin and infinity of $\mathbb{C}$. The insertion points $\Sigma_q$ are now mapped to 
\begin{align}\label{eqn:uniformisation}
    w^{(k)}_1 =\bigg(\frac{z^{(k)}_1-u}{z^{(k)}_1-v}\bigg)^{\frac{1}{q}}\,,\quad \bar{w}^{(k)}_1=\bigg(\frac{\bar{z}^{(k)}_1-\bar{u}}{\bar{z}^{(k)}_1-v}\bigg)^{\frac{1}{q}}\,;\nn
    w^{(k)}_2 =\Bigg(\frac{\frac{1}{\bar{z}^{(k)}_1}-u}{\frac{1}{\bar{z}^{(k)}_1}-v}\Bigg)^{\frac{1}{q}}\,,\quad \bar{w}^{(k)}_2=\Bigg(\frac{\frac{1}{z^{(k)}_1}-\bar{u}}{\frac{1}{z^{(k)}_1}-v}\Bigg)^{\frac{1}{q}}\,,
\end{align}
on $\mathbb{C}$.
In the uniformization map, the primary operator $\mathcal{O}$ of dimension $(h,h)$ transforms as
\begin{align}\label{eqn:primary}
    \mathcal{O}(z, \bar{z})=\bigg(\frac{d w}{d z} \frac{d \bar{w}}{d \bar{z}}\bigg)^h \mathcal{O}(w,\bar{w})\,.
\end{align}
The $2q$-point correlation function appearing in \eqref{eqn:ee2} transforms to the following $2q$-point correlation function on $\mathbb{C}$
\begin{align}
\langle \prod_{k=0}^{q-1}\mathcal{O}(z^{(k)}_1, \bar{z}^{(k)}_1) \mathcal{O}^\dagger(z^{(k)}_2, \bar{z}^{(k)}_2)\rangle_{{{\Sigma}}_q}    =\mathcal{J}^q_{\mathcal{O}^\dagger} \mathcal{J}^q_{\mathcal{O}}\langle \prod_{k=0}^{q-1}\mathcal{O}(w^{(k)}_1, \bar{w}^{(k)}_1) \mathcal{O}(w^{(k)}_2, \bar{w}^{(k)}_2)\rangle_{\mathbb{C}}\,,
\end{align}
where $\mathcal{J}^q_{\mathcal{O}^\dagger}$ is the Jacobian factor coming from the transformation \eqref{eqn:primary}
\begin{align}\label{transformation}
  \mathcal{J}^q_{\mathcal{O}} &= \prod_{k=0}^{q-1}\Bigg(\frac{d w^{(k)}_1}{d z^{(k)}_1}\frac{d \bar{w}^{(k)}_1}{d z^{(k)}_1}\Bigg)^h\,,\nn
  \mathcal{J}^q_{\mathcal{O}^\dagger}  &=\prod_{k=0}^{q-1} \Bigg(\frac{d w^{(k)}_2}{d \bar{z}^{(k)}_1}\frac{d \bar{w}^{(k)}_2}{d z^{(k)}_1}\Bigg)^h\,.
\end{align}

Using this map, we can compute the entanglement entropy as: 
\begin{align}\label{eqn:ee3}
  & \delta S_{EE}(\tau)\nn &=\underbrace{\lim_{q\rightarrow1}  \frac{1}{1-q}\log \bigg[\frac{\mathcal{J}^q_{\mathcal{O}^{\dagger}} \mathcal{J}^q_{\mathcal{O}}}{\big(\tilde{\mathcal{J}}^1_{\mathcal{O}^\dagger} \tilde{\mathcal{J}}^1_{\mathcal{O}}\big)^q}\bigg]}_{S^{uni}}
  +\underbrace{\lim_{q\rightarrow1}  \frac{1}{1-q} \log \bigg[\frac{\langle \prod_{k=0}^{q-1}\mathcal{O}(w^{(k)}_1, \bar{w}^{(k)}_1) \mathcal{O}(w^{(k)}_2, \bar{w}^{(k)}_2)\rangle_{\mathbb{C}}}{{\langle \mathcal{O}(w^{(0)}_1, \bar{w}^{(0)}_1) \mathcal{O}(w^{(0)}_2, \bar{w}^{(0)}_2)\rangle}^q _\mathbb{C}}\bigg]}_{S^{dyn}}\,,
\end{align}
where
\begin{equation}
    \tilde{\mathcal{J}}^1_{\mathcal{O}^\dagger} \tilde{\mathcal{J}}^1_{\mathcal{O}} \equiv \frac{\langle \mathcal{O}(z^{(0)}_1, \bar{z}^{(0)}_1) \mathcal{O}^\dagger(z^{(0)}_2, \bar{z}^{(0)}_2)\rangle_\mathbb{C}}{\langle \mathcal{O}(w^{(0)}_1, \bar{w}^{(0)}_1) \mathcal{O}(w^{(0)}_2, \bar{w}^{(0)}_2)\rangle_\mathbb{C}}\,.
\end{equation}
It follows that the entanglement entropy can have two contributions:
\begin{enumerate}[label={(\alph*)}]
    \item $S^{uni}:$ the part of the entropy that is fixed by kinematics and is universal.
    \item $S^{dyn}:$ the part of the entropy which depends on the dynamics of the CFT and the specific form of the correlation function.
\end{enumerate}
The universal part of the regularised entanglement entropy can be written in terms of the insertion points  \cite{Agon:2020fqs}
\begin{align}\label{eqn:suni}
    S^{uni}=2h\left(2+\frac{1}{2}\left(\frac{w_2+w_1}{w_2-w_1}\right)\log\frac{w_1}{w_2}+\frac{1}{2}\left(\frac{\bar{w}_2+\bar{w}_1}{\bar{w}_2-\bar{w}_1}\right)\log\frac{\bar{w}_1}{\bar{w}_2}\right)\,.
\end{align}
The insertion points can be obtained using the m\"{o}bius transformation (see Appendix \ref{appendix:mobius}) and finally setting $q=1$ in \eqref{eqn:uniformisation} which results in the following 
 \begin{eqnarray}\label{eqn:insertion_points}
  &&  \omega_1 =\frac{z_n(z=0) - e^{2\pi ix}}{z_n(z=0) -1}, \quad  \omega_{2} = \frac{z_n(z=\infty) - e^{2\pi i x}}{z_n(z=\infty) -1} \nonumber\,, \\
   &&  \bar{\omega}_1 =\frac{z_n(z=0) - e^{-2\pi ix}}{z_n(z=0) -1}, \quad  \bar{\omega}_{2} = \frac{z_n(z=\infty) - e^{-2\pi i x}}{z_n(z=\infty) -1}\,,
\end{eqnarray}
where $z_n(z=0)$ and $z_n(z=\infty)$ for different phases  are given in \eqref{eqn:mob_hp}, \eqref{eqn:mob_nhp} and \eqref{eqn:mob_pb}.

\begin{figure}[!h]
\centering
\begin{tikzpicture}[scale=3]

\draw[thick] (-1.2,0)--(1.2,0);
\draw[thick] (0,-1.2)--(0,1.2);
\draw[thick] (-0.85,-0.85)--(0.85,0.85);
\draw[thick] (-0.85,0.85)--(0.85,-0.85);

\filldraw[red] (0.98,0.19) circle (0.5pt) node[anchor=west,text=black]{$w^{(0)}_1$};
\filldraw[red] (0.78,0.15) circle (0.5pt) node[anchor=east,text=black]{$w^{(0)}_2$};

\filldraw[blue] (0.55,0.83) circle (0.5pt) node[anchor=west,text=black]{$w^{(1)}_1$};
\filldraw[blue] (0.44,0.66) circle (0.5pt) node[anchor=east,text=black]{$w^{(1)}_2$};

\filldraw[red] (-0.19,0.98) circle (0.5pt)node[anchor=south,text=black]{$w^{(2)}_1$};
\filldraw[red] (-0.15,0.78) circle (0.5pt) node[anchor=north,text=black]{$w^{(2)}_2$};

\filldraw[red] (-0.83,0.55) circle (0.5pt)node[anchor=north east,text=black]{$w^{(3)}_1$};
\filldraw[red] (-0.66,0.44) circle (0.5pt) node[anchor=north east,text=black]{$w^{(3)}_2$};

\filldraw[red] (-0.98,-0.19) circle (0.5pt)node[anchor=north,text=black]{$w^{(4)}_1$};
\filldraw[red] (-0.78,-0.15) circle (0.5pt) node[anchor=north west,text=black]{$w^{(4)}_2$};

\filldraw[red] (-0.55,-0.83) circle (0.5pt)node[anchor=north west,text=black]{$w^{(5)}_1$};
\filldraw[red] (-0.44,-0.66) circle (0.5pt) node[anchor=south west,text=black]{$w^{(5)}_2$};

\filldraw[blue] (0.19,-0.98) circle (0.5pt)node[anchor=north,text=black]{$w^{(6)}_1$};
\filldraw[blue] (0.15,-0.78) circle (0.5pt) node[anchor=south,text=black]{$w^{(6)}_2$};

\filldraw[red] (0.83,-0.55) circle (0.5pt)node[anchor=north west,text=black]{$w^{(7)}_1$};
\filldraw[red] (0.66,-0.44) circle (0.5pt) node[anchor=south,text=black]{$w^{(7)}_2$};
\end{tikzpicture}
\caption{Sub-leading order correction to $S^{dyn}$ in the short distance expansion and the uniformisation map for $q=8$. The operators contracted on the same wedge are marked red and the four point function of operators on two different wedges are marked blue.}\label{fig:uniformmap}
\end{figure}
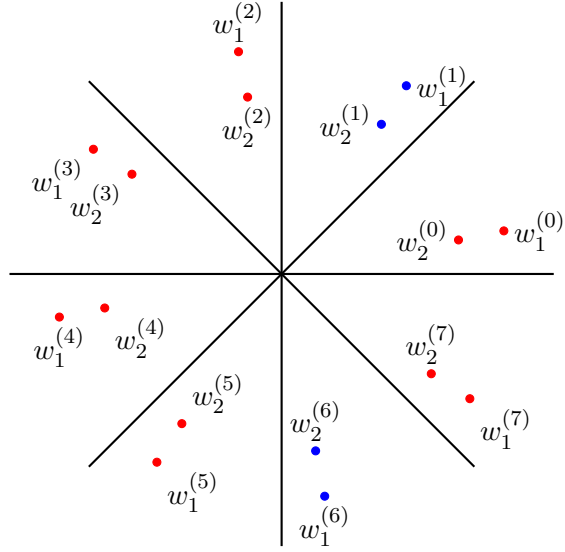

Next, we evaluate the dynamical part $S^{dyn}$. It is difficult to compute the exact expression of the $2q$-point correlator in \eqref{eqn:ee3}. We use the short distance expansion method following \cite{Sarosi:2016oks, Agon:2020fqs}. In this approximation, the distance between the operators $\mathcal{O}$ and $\mathcal{O}^\dagger$ in a given wedge, specified by the value of $k$, is small and the leading contribution to the $2q$-point correlation function is given by the product of the $q$ 2-point functions. In the OPE expansion, this corresponds to taking the identity contribution. Furthermore, the first sub-leading correction comes from the four point correlation function of the operators in which a pair of operators are placed at one of the wedges and another pair in another wedge. The remaining operators are contracted as pairs on the same wedge (see Figure \ref{fig:uniformmap}).

This can be evaluated using the methods developed in \cite{Calabrese:2010he}. We provide  the final expression below:
\begin{align}
    \log \bigg[\frac{\langle \prod_{k=0}^{q-1}\mathcal{O}(w^{(k)}_1, \bar{w}^{(k)}_1) \mathcal{O}(w^{(k)}_2, \bar{w}^{(k)}_2)\rangle_{\mathbb{C}}}{{\langle \mathcal{O}(w^{(0)}_1, \bar{w}^{(0)}_1) \mathcal{O}(w^{(0)}_2, \bar{w}^{(0)}_2)\rangle}^q _\mathbb{C}}\bigg]\approx (q-1)\frac{\Gamma(\frac{3}{2})\Gamma(4h+1)}{\Gamma(4h+\frac{3}{2})}\bigg(\frac{w_1-w_2}{w_1+w_2}\cdot \frac{\bar{w}_1-\bar{w}_2}{\bar{w}_1+\bar{w}_2}\bigg)^{4h}\,.
\end{align}
Hence, the dynamical part of the entanglement entropy \eqref{eqn:ee3}, in the short distance approximation, can be written as
\begin{align}\label{eqn:sdyn}
    S^{dyn}&= -\frac{\Gamma(\frac{3}{2})\Gamma(4h+1)}{\Gamma(4h+\frac{3}{2})}\bigg(\frac{w_1-w_2}{w_1+w_2} \cdot\frac{\bar{w}_1-\bar{w}_2}{\bar{w}_1+\bar{w}_2}\bigg)^{4h}+\cdots\,.
    \end{align}
 The ellipsis in \eqref{eqn:sdyn} refers to the higher order terms in the small $x$ limit. Combining \eqref{eqn:suni} and \eqref{eqn:sdyn} we can obtain the entanglement entropy 
\begin{align}\label{eqn:cftfull}
    S_{tot}=S^{uni}+S^{dyn}\,.
\end{align}
Finally, we can analytically continue to the Lorentzian time using Wick rotation $\tau \rightarrow i t$ and obtain the Lorentzian entanglement entropy. This completes the derivation of the entanglement entropy using the replica method. In the following subsections we compute \eqref{eqn:cftfull} for various phases in \eqref{eqn:3phases}.

\subsection{Entanglement entropy in different phases}\label{sec:ee_3phases}
In this subsection, we compute the insertion points for different phases and then evaluate the total entanglement entropy\eqref{eqn:cftfull}.
\paragraph{Heating and non-heating phase:} ~\\
The insertion points for the heating and non-heating phases are given in \eqref{eqn:insertion_points}, together with \eqref{eqn:mob_hp}  \eqref{eqn:mob_nhp}, under the analytic continuation $\tau\rightarrow i t, T\rightarrow iT$ .

\begin{figure}[!h]
   \centering          
   \includegraphics[width=0.5\textwidth]{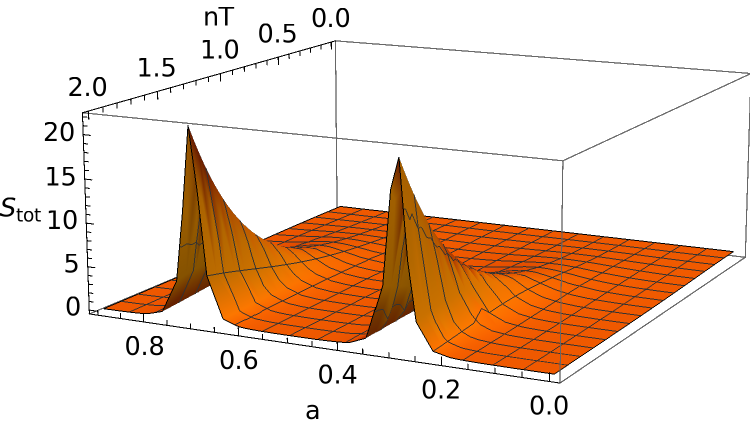} 
   \includegraphics[width=0.45\textwidth]{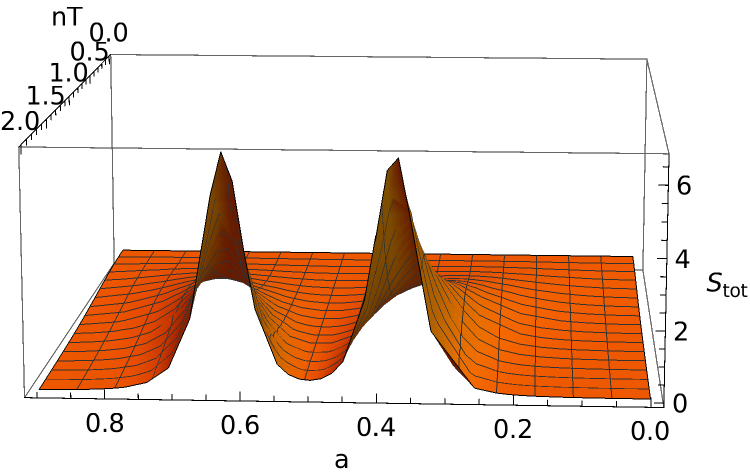}
   \caption{Leading order behavior of $S_{tot}= S^{uni}+S^{dyn}$ for \eqref{eqn:stot_full} in the small $x$ limit, as a function of $nT$ for $ h=3, t=0$ and $x=0.1$ in the entangling region \eqref{eqn:interval_a}.  {\it Left:} Heating phase for $\alpha=0.4, \beta=0.6$ has two peaks around $a=0.3 $ and $a=0.7$ {\it Right:} Phase boundary for $\alpha=1$ has two peaks around $a=0.4$ and $a=0.6$. }
   \label{fig:ee_plot3d}
\end{figure}

Substituting these in \eqref{eqn:suni} and \eqref{eqn:sdyn}, we obtain the following expression for the evolution of the entanglement entropy in the heating and the non- heating phase. 
\begin{align}\label{eqn:stot_full}
  S^{uni} &= \frac{2h }{\delta ^2 \mathcal{S}_{\delta  n T}}\bigg(2 \mathcal{K}-\mathcal{M}\tan ^{-1}\left(\frac{\mathcal{K}}{\mathcal{M}}\right)-\mathcal{P}\tan ^{-1}\left(\frac{\mathcal{K}}{\mathcal{P}}\right)\bigg)\,,\nn
  S^{dyn}&= -\frac{e^{8 i \pi  h}\Gamma(4h+1)\sqrt{\pi}}{2\Gamma(4h+\frac{3}{2})}\bigg(\frac{\mathcal{K}^2}{\mathcal{M}\mathcal{P}}\bigg)^{4h}+\cdots\,,
\end{align}
where
\begin{align}
\mathcal{K}&=\delta ^2  \mathcal{S}_{\delta  n T}\,\\
\mathcal{M}&={ \csc(\pi  x)} \bigg(\left(\alpha ^2+4 \beta ^2\right) \cos (\pi  x)+4\alpha\beta \cos (t+\pi x)+ \delta \mathcal{C}_{\delta  n T} \bigg(\delta  \cos{(\pi x)}  \mathcal{C}_{\delta  n T}+4  \beta  \sin(t+\pi x)\bigg)\bigg)\,,\nn 
\mathcal{P}& \notag ={ \csc(\pi  x)} \bigg(\left(\alpha ^2+4 \beta ^2\right) \cos (\pi  x)+4\alpha\beta \cos (t-\pi x)+ \delta \mathcal{C}_{\delta  n T} \bigg(\delta  \cos{(\pi x)}  \mathcal{C}_{\delta  n T}+4  \beta  \sin(t-\pi x)\bigg)\bigg)\, ,
\end{align}
with
 \begin{align}
\mathcal{S}_\rho &=
\begin{cases}
\csc^2 \frac{\rho}{2}, & \text{non-heating phase},\\
\csch^2 \frac{\rho}{2}, & \text{heating phase},
\end{cases}
\\[1ex]
\mathcal{C}_\rho &=
\begin{cases}
\cot \frac{\rho}{2}, & \text{non-heating phase},\\
\coth \frac{\rho}{2}, & \text{heating phase}.
\end{cases}
\end{align}
The quantity $\delta$ is defined in \eqref{eqn:delta_def} for different phases. It turns out that the entanglement entropy oscillates as a function of $t$ for all $t > n T$. This is expected since we have evolved the initial state by the unperturbed CFT Hamiltonian $H_0$.
\begin{figure}[!h]
   \centering        \includegraphics[width=0.52\textwidth]{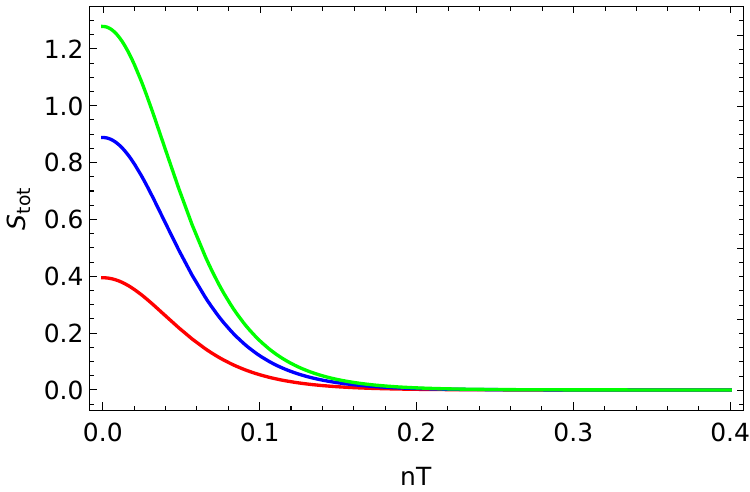}\includegraphics[width=0.51\textwidth]{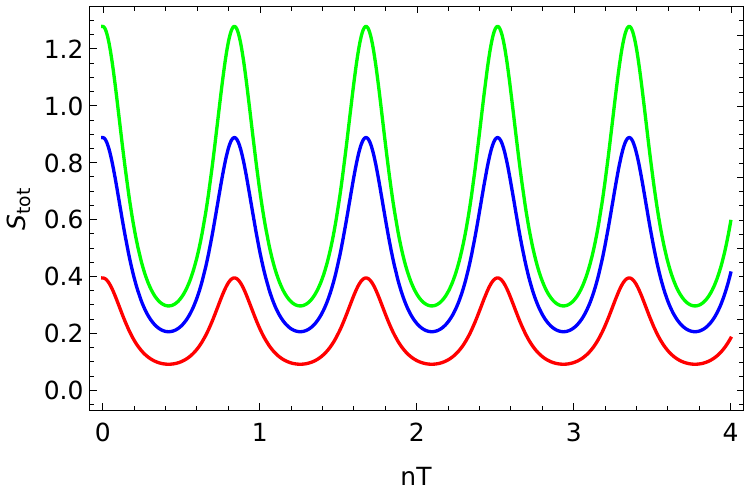} 
   \\[0.5cm]
   \includegraphics[width=0.6\textwidth]{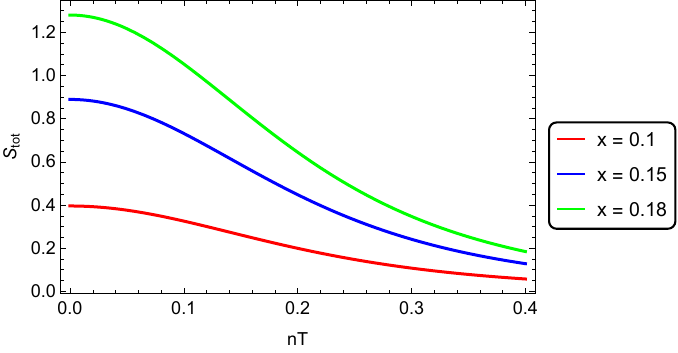}
   \caption{Leading order behavior of $S_{tot}= S^{uni}+S^{dyn}$ for \eqref{eqn:stot_full} in the small $x$ limit, as a function of $nT$ for $ h=3, t=0$ and  different values of the interval width $x=0.1, 0.15,0.18$. Here the origin of the interval is $a=0$. {\it Top left:} Heating phase with $\alpha=1.4$ and $ \beta=8$. {\it Top right:} Non-heating phase with  $\alpha=8$ and $ \beta=1.4$. {\it Bottom}: Phase boundary with $\alpha=3.2$.  }
   \label{fig:ee_plot2d}
   \end{figure}
\paragraph{Phase boundary:} ~\\\
The expression for the evolution of the entanglement entropy in the phase boundary reads
\begin{align}\label{phase3}
    S^{uni} &= h \csc (\pi  x)\bigg[4 \sin(\pi x)-g(x) \tan ^{-1}\left(\frac{2\sin (\pi x)}{g(x)}\right)-g(-x)\tan ^{-1}\left(\frac{2\sin (\pi x)}{g(-x)}\right)\bigg]\,,\nn
    S^{dyn}&=- e^{8 i \pi  h}\frac{2h\sqrt{\pi}\Gamma(4h)}{\Gamma(4h+\frac{3}{2})}\bigg(\frac{4\sin^2(\pi x)}{g(x)g(-x)}\bigg)^{4h}+\cdots\,,
    \end{align}
    where
    \begin{align}
        g(x) &=\left((n T \alpha) ^2+2\right) \cos (\pi  x)+n T\alpha  \bigg(n T\alpha  \cos (t-\pi  x)+2  \sin (t-\pi  x)\bigg)\,.
    \end{align}
It can be shown that the expression above for the phase boundary is continuous from that in both the heating and the non-heating phase found in \eqref{eqn:stot_full}.    
Now let us change the interval endpoints from \eqref{eqn:interval} to the following
\begin{align}\label{eqn:interval_a}
    u=e^{2\pi i (x+a)}, \quad v=e^{2\pi i a}\,, a \in (0,1)
\end{align}
We can evaluate the full entropy $S_{tot}$ with the modified interval \eqref{eqn:interval_a}. This concludes the evaluation of semiclassical entanglement entropy for various drive parameters corresponding to the three phases. Now we put $t=0$ and vary them with stroboscopic time $nT$ to note different behaviors for different phases. The behavior of $S_{tot}$ with $nT$ for different phases are given in Figure \ref{fig:ee_plot2d} and \ref{fig:ee_plot3d}. 
 A summary of these plots is provided below.
\begin{enumerate}
\item \underline{Heating phase}: We consider $\alpha=1.4$ and $\beta=8$ to note damping behavior for $a=0$ (fig. \ref{fig:ee_plot2d}) and two peaks around $a=0.3$ and $a=0.7$ ( fig \ref{fig:ee_plot3d}). 
This can be explained as follows. In order to observe the growing entanglement entropy in the heating phase, the interval should contain either of the energy peak . In intervals without an energy peak or both peaks, the entanglement entropy should saturate to a $O(1)$ value \cite{Fan:2019upv}. Due to the short distance approximation, we note two peaks of entanglement entropy around $a=0.3$ and $0.7$ respectively (fig. \ref{fig:ee_plot3d}).

\item \underline{Non-heating phase}: We set $\alpha=8$ and $\beta=1.4$ to observe the oscillatory behavior of $S_{tot}$ (fig. \ref{fig:ee_plot2d}). The oscillation period increases near the phase boundary.

\item \underline{Phase Boundary}: We consider $\alpha=3.2$ with $a=0$ to note a slow decay in the entanglement entropy (fig. \ref{fig:ee_plot2d}). For non-zero $a$, there are two peaks in the entanglement entropy at $a=0.4$ and $0.6$, respectively (fig. \ref{fig:ee_plot3d}).
\end{enumerate}

\subsubsection{An example of a holomorphic drive}
\label{subsec: CFT alpha=0}
In this subsection, we evaluate the time evolution of the entanglement entropy of a state which has been prepared by a $n$-cycle drive. The corresponding effective Hamiltonian is holomorphic with $\alpha=0$, i.e.,
\[  H_{eff}=\beta(L_1+L_{-1}).\]
This is of special interest, as we will focus on the holographic dual of this state in the next section. The prepared state can be simplified using BCH formula \eqref{eqn:normalorder}. The simplified expression of this state has been found in \eqref{eq: prepared state cft normal order} and \eqref{eq: non-heating parameter CFT}. For convenience of readers, we also provide it below.
\begin{align}\label{eq: prepared state cft normal order main}
    |\hat{\psi}\rangle = e^{h\log[A_0^R]} \sum_{q=0}^\infty \frac{(A_+^R)^q}{q!} L_{-1}^q |h,h \rangle, 
\end{align}
with the parameters
\begin{equation}\label{eq: non-heating parameter CFT main}
    A_{+}^R=-i \tanh \sigma,\ A_{-}^L=(A_{+}^R)^*,\ A_0^L=A_0^R= (\cosh \sigma)^{-2}.
\end{equation}
In this case, we have the following insertion points 
\begin{align}\label{operatorpositions}
    w_1&=\frac{e^{- i t}-i e^{2\pi i x}\coth{(n T\beta)}}{e^{-i t}-i \coth{(n T\beta)}}, \qquad  \bar{w}_1=e^{-2\pi i x},\nn
 w_2&=\frac{ie^{- i t}\coth{(n T\beta)}+ e^{2\pi i x}}{ie^{- i t}\coth{(n T\beta})+1}, \qquad  \bar{w}_2=1\,.   
\end{align}
This results in the universal term 
\begin{align}\label{eqn:cftuni}
    S^{uni}&=2h \left(2-\pi x \cot(\pi x)- A  \cot ^{-1}A \right)\,,
\end{align}
where
\begin{align}
   A &=\cot (\pi  x) \cosh (2n \beta  T)+\sinh (2 n \beta  T)\csc(\pi x) \sin(t+\pi  x)\,.
\end{align}
Note that the expression in \eqref{eqn:cftuni} is valid for any $x$ and $n T \beta$. The dynamical part of the entanglement entropy \eqref{eqn:sdyn} in the short distance approximation can be written as
\begin{align}\label{eqn:sdyn2}
    S^{dyn}&= -\frac{2h\ \sqrt{\pi}\Gamma(4h)}{\Gamma(4h+\frac{3}{2})}\bigg(\frac{(\pi x)^2}{\sin t \sinh (2 n\beta  T)+\cosh (2n \beta  T)}\bigg)^{4h}+\cdots.
    \end{align}
Our goal would be to obtain \eqref{eqn:cftuni} and\eqref{eqn:sdyn2} from holography. The expectation is that $S^{uni}$ obtained from the CFT should match the correction to the minimal area term due to the change in the dual bulk geometry. Also, $S^{dyn}$ obtained from the CFT side should agree with the second order correction to the bulk entanglement entropy. In \eqref{eq: final holo ee}, we will see that this is indeed the case.


\section{Entanglement entropy in the bulk}\label{3}
\label{sec: bulk computation}
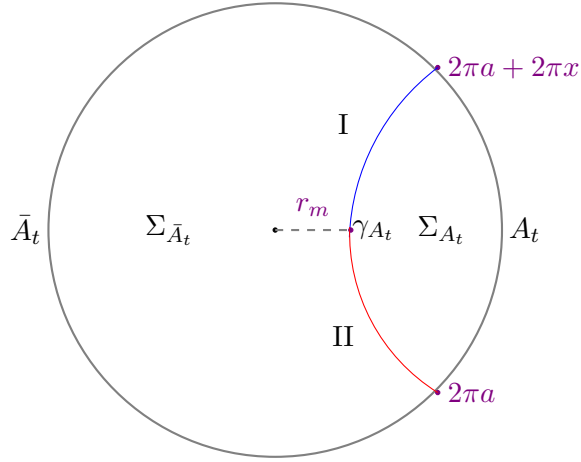
\begin{figure}[h]
\begin{center}
\begin{tikzpicture}[rotate=90,transform shape]
\coordinate (A) at (0,-0.5);
\node[xshift= 0.3 cm,rotate =-90] at (A) {\color{violet}{$ r_m $}};
\coordinate (B) at (-1.4,-0.9);
\node [rotate=-90]at (B) {II};
\coordinate (C) at (1.4,-0.9);
\node[rotate=-90] at (C) {I};
\coordinate (D) at (0,-1.3);
\node[rotate=-90] at (D) {$\gamma_{A_t}$};
\coordinate (E) at (0,-2.2);
\node[rotate=-90] at (E) {$\Sigma_{A_t}$};
\coordinate (F) at (0,1.4);
\node[rotate=-90] at (F) {$\Sigma_{\bar{A}_t}$};
\coordinate (G) at (0,3.3);
\node[rotate=-90] at (G) {$\bar{A}_t$};
\coordinate (H) at (0,-3.3);
\node[rotate=-90] at (H) {$A_t$};
\draw[color=gray,thick] (0,0) circle [radius=3];
\filldraw [violet] (0,-1) circle (0.8 pt) node{};
\filldraw [violet] (2.15, -2.15) circle (0.8 pt) node[rotate=-90,below][anchor=west]{$ 2\pi a+ 2\pi x$};
\filldraw [violet] (-2.15, -2.15) circle (0.8 pt) node[rotate=-90,below][anchor=west]{$ 2 \pi a $};
\filldraw [black] (0,0) circle (0.8 pt) node{};
\draw[ dashed, thick, gray!100] (0,0)--(0,-1);
\draw[blue] (2.15,-2.15) arc (37:88:2.94);
\draw[red] (-2.15,-2.15) arc (148:88:2.48);
\end{tikzpicture}
\end{center}
\caption{We show a constant time slice at time $t$ of $AdS_3$. $\gamma_{A_t}$ denotes the minimal length of subsystem $A_t$ in the boundary. $\gamma_{A_t}$ splits the bulk into right(left) wedge denoted by $\Sigma_{A_t}(\Sigma_{\bar{A}_t})$. $\gamma_{A_t}$ consists of :  Branch I, where slope $\varphi^\prime(r)>0$ and Branch II with $\varphi^\prime(r) < 0$. Our excited state \eqref{eq: primary state bulk} breaks  the angular symmetry, hence we need to evaluate the minimal area for these two branches separately. We have aligned the origin of the interval to the origin of the angular coordinate $\varphi$.}
\label{fig: Rindler}
\end{figure}

In this section, we compute the time evolution of the holographic entanglement entropy of the dual CFT state $| \hat{\psi}\rangle$ in \eqref{eq: prepared state cft normal order main}. We consider an interval in the boundary CFT of size $(0, 2\pi x)$ at constant time slice at $t$. We denote this interval as the subsystem $A_t$. We would like to find the holographic entanglement entropy between $A_t$ and its complimentary subsystem $\bar{A}_t$ (see fig. \ref{fig: Rindler}) at the order $G_N^0$. Note that in the CFT computation we have assumed that the interval or subsystem is located at $t=0$. This is fine as long as the time is a Killing vector in the bulk geometry. The back-recation, due to the excited state we considered here, induces a time-dependent perturbation in the metric in our case, but this does not affect the $G_N^0$ order results. For convenience, we will explicitly compute the entropy for the interval at $(0, 2 \pi x)$, with the drive parameter $\alpha=0$. At the end of this section, we shall generalize for the interval $(2 \pi a, 2 \pi a+ 2\pi x)$ and comment about the case with an arbitrary $\alpha$.

We start with the gravitational theory of a minimally coupled scalar field $\phi$ in $AdS_3$.
\begin{align}
S=\int d^3 x \sqrt{-g}\left(\frac{1}{16\pi G_N}(R+2)-\frac{1}{2}(\nabla \phi)^2-\frac{1}{2}M^2 \phi^2\right),
\end{align}
where $M$ is the mass of the scalar field, which is related to the conformal dimension $(h,h)$ of the boundary operator as follows
\begin{align}
M^2= 4h(h-1)\,.
\end{align}
The $AdS_3$ metric can be written in global coordinates
\begin{align}
    & ds^2= -(1+r^2) dt^2+ \frac{1}{(1+r^2)} dr^2+ r^2 d \varphi,\,\,\varphi\rightarrow \varphi+ 2 \pi.
\end{align}
We denote $|0_G \rangle$ as the vacuum of the bulk in the global coordinate, which is dual to the vacuum in the boundary CFT $|0,0\rangle $. The time $t$ is identified with CFT Lorentzian time, $t$. We solve the Klein-Gordon equation
\begin{align}\label{eqn:kg}
    \left( \Box-4h(h-1) \right)\phi=0\,.
\end{align}
to obtain the mode expansion of $\phi$ as follows.
\be\label{modexp}
\phi(t,r,\varphi) = \sum_{m,n}^\infty \le( a_{m,n} e^{-i \Omega_{m,n} t}e^{-im\varphi} f_{m,n}(r) + a^{\dagger}_{m,n} e^{i \Omega_{m,n} t}e^{im\phi} f^{*}_{m,n}(r)\ri) \, .
\ee
The solutions, regular at $r=0$ are given by 
\begin{align}
\notag f_{m, n}(r)  =& N_{m, n}  r^m ( 1+ r^2)^{\frac{\Omega_{m, n} }{2}}
 {}_2F_{1}\Big( \frac{1}{2} ( m - 2h +2 + \Omega_{m, n} ), 
\frac{1}{2} ( m + 2h + \Omega_{m, n} ) , 1+m; -r^2 \Big) , \\
\text{for}& \;m \in \mathbb{Z} \,.
\end{align}
Demanding the finiteness of these functions  at $r\rightarrow\infty$ results in the quantization condition 
\begin{equation}
\Omega_{m, n} = 2h + |m| + 2 n, \qquad n \in \mathbb{Z}_+\,.
\end{equation}
The normalization constants $N_{m, n}$ are fixed using the standard Klein-Gordon inner product which is given by
\begin{align}
& \notag 2\Omega_{m,n}\int_0^{\infty} dr \int_0^{2\pi} d\varphi \sqrt{-g} g^{tt}(r) f_{m,n}(r) f_{m',n'}^*(r) = \delta_{n,n'} \delta_{m,m'}\,.
\end{align}
For the holomorphic sector, we will need to consider the modes with $n=0$ and arbitrary $m$, i.e., the wavefunctions of the form $f_{m,0}$. The normalization constants for these modes are given by
\begin{align}\label{eq: normalization const bulk}
    N_{m,0}=\sqrt{\frac{\Gamma (2 h+m)}{2 \pi\Gamma (2 h) \Gamma (m+1)}}.
\end{align}
To construct the dual state in the bulk, we simply need to lift the global conformal generators to the isometry generators in the bulk geometry. The bulk dual state of the primary and holomorphic descendants with non-zero angular momentum in the boundary CFT follows the dictionary provided in Table \ref{eq: cft bulk dictionary}.
{\small
\begin{table}[h!]
\centering
\begin{tabular}{ |c|c|c| } 
 \hline
State & Operator & Wave function \\ 
 \hline
 Primary: $|h,h\rangle \equiv |\psi_{0,0}\rangle$ & $a_{0,0}^\dagger|0\rangle_{G} \equiv |\Psi_{0,0}\rangle $& $f_{0,0}(r) e^{-2iht}$ \\ 
 \hline
  Descendants: $L_{-1}^m|h,h\rangle \equiv |\psi_{m,0}\rangle$ & $a_{m,0}^\dagger |0\rangle_G \equiv |\Psi_{m,0}\rangle$ & $ f_{m,0}(r) e^{-i \Omega_{m,0}t}e^{-i m \varphi}$ \\ 
 &$\qquad \qquad\quad=\notag \frac{1}{\mathcal{N}_{m,0}}\mathcal{L}_{-1}^m|\Psi_{0,0}\rangle$& \\
 \hline
\end{tabular}
\caption{States, operators and corresponding wave functions}
\label{eq: cft bulk dictionary}
\end{table}
}
Note that $\mathcal{L}_{-1,0,1}$ denotes the $AdS$ isometry generators and $\mathcal{N}_{m,0}^2= \langle h,h|L_{1}^m L_{-1}^m|h,h\rangle=\frac{\Gamma(2h+m) m!}{\Gamma(2h)}$ is the norm of the $m$-th level descendant $|\psi_{m,0} \rangle$ in CFT. The normalized primary state can be obtained as \cite{Maldacena:1998bw}
\begin{align}
    |\Psi_{0,0}\rangle=\frac{1}{\sqrt{2 \pi }}\frac{e^{-2h i t}}{ \left(r^2+1\right)^h}.
\end{align}
Using this dictionary, we can construct a bulk dual of the boundary state $|\hat{\psi} \rangle$ in \eqref{eq: prepared state cft normal order main}.
\begin{align}\label{eq: primary state bulk}
    \notag |\hat{\Psi}\rangle & = e^{A_+^R \mathcal{L}_{-1}} e^{\mathcal{L}_0\log[A_0^R]} e^{A_-^R\mathcal{L}_1}|\Psi_{0,0}\rangle\\
    & = e^{h\log[A_0^R]} \sum_{q=0}^\infty \frac{(A_+^R)^q}{q!} \mathcal{N}_{q,0} a_{q,0}^\dagger |\Psi_{0,0} \rangle \,.
\end{align}
Note that the dictionary in Table \ref{eq: cft bulk dictionary} ensures that this state has the same norm as that of the dual state in the boundary.

\subsection{Backreacted geometry}

The dual state $|\hat{\Psi}\rangle$ in \eqref{eq: primary state bulk} is a linear superposition of primary and all global holomorphic descendants with non-zero angular momentum. This will leave a  backreaction on pure $AdS_3$ for $G_N \neq 0$. We solve Einstein's equation and find the backreacted metric $g_{\mu\nu}$ from the Einstein's equation in the $G_N \rightarrow 0$ limit.
\begin{align}\label{eq: Einstein}
R_{\mu \nu}-\frac{1}{2}g_{\mu \nu}-g_{\mu \nu} =G_N \langle \hat{\Psi} | T_{\mu\nu}|\hat{\Psi}\rangle\,,
\end{align}
where the metric can be expanded in $G_N$ as follows
\begin{align}
g_{\mu \nu} =g^{(0)}_{\mu \nu} + G_N \delta g^{(1)}_{\mu \nu}+ \cdots \,.
\end{align}
At first order in $G_N$, Einstein's equation is linear with a suitable gauge choice. The normal ordered stress tensor for the scalar field $\phi$ is given by
\begin{align}\label{eqn:stress_tensor}
T_{\mu \nu}=:\partial_{\mu}\phi \partial_{\nu}\phi-\frac{1}{2}g^{(0)}_{\mu \nu}\left((\nabla \phi)^2+m^2\phi^2\right):\,.
\end{align}
 We can compute the expectation value of $T_{\mu\nu}$ for the bulk state $|\hat{\Psi}\rangle$ in \eqref{eq: primary state bulk} using the BCH formula in \eqref{eq: observable cft} and the dictionary in Table \ref{eq: cft bulk dictionary}.
\begin{align}\label{eq: stress tensor bulk}
   & \langle \hat{\Psi}|T_{\mu\nu}|\hat{\Psi}\rangle\\
& \nonumber = e^{h\log A_0^L }e^{h\log A_0^R}\sum_{m,n=0}^\infty \frac{(A_-^L)^m}{m!}  \frac{(A_+^R)^n}{n!} \mathcal{N}_{m,0}\mathcal{N}_{n,0} \langle 0_G | a_{m,0} T_{\mu\nu} a_{n,0}^\dagger |0_G \rangle\,.
\end{align}
$\mathcal{N}_{m,0}$ is the norm of the state $|\psi_{m,0} \rangle$ in the boundary. The expressions of parameters $A_-^L, A_+^R, A_0^L, A_0^R$ are given in \eqref{eqn:normalorder}. For $\alpha=0$ these parameters become (see \eqref{eq: non-heating parameter CFT})
\begin{equation}\label{eq: non-heating parameter CFT}
    A_{+}^R=-i \tanh \sigma,\ A_{-}^L=(A_{+}^R)^*,\ A_0^L=A_0^R=(\cosh \sigma)^{-2}\,,
\end{equation}
where  $\sigma$ is defined in \eqref{eq: def sigma}. For our choice of parameters in \eqref{eq: non-heating parameter CFT} 
\[\sigma= n T \beta=n T \gamma= \frac{n T \delta}{2}. \]
We have defined the $sl(2,\mathbb{R})$ invariant, $\delta$ in \eqref{eqn:delta_def}. Because $\alpha=0$, this corresponds to the heating phase. We also use the following notation for brevity
\begin{align} \label{eq: Cm general def}
    \frac{(A_-^L)^m}{m!}= C_m^*,\, \frac{(A_+^R)^n}{n!}= C_n\,.
\end{align}
We rewrite \eqref{eq: stress tensor bulk} as
\begin{align} \label{eq: stress tensor formula}
   \langle \hat{\Psi}|T_{\mu\nu}|\hat{\Psi}\rangle & = (\cosh \sig )^{-4 h}\sum_{m,n=0}^\infty \mathcal{N}_{m,0}\mathcal{N}_{n,0} C_m^* C_n \langle 0_G| a_{m,0} T_{\mu\nu} a_{n,0}^\dagger |0_G \rangle\,.
\end{align}
We will spend a few words on how we will proceed to compute the backreaction of $|\hat{\Psi}\rangle$. First, we need to find the backreacted geometry for the bulk dual state $| \hat{\Psi}^{m,n} \rangle $ corresponding to the superposed state $|\hat{\psi}^{m,n} \rangle$ in the boundary, which is a linear combination of holomorphic descendants at arbitrary $m$-th and $n$-th level in the boundary for $m \neq n$.  The correspondence can be written as
\begin{align}\label{eq: cft bulk dictionary superposed}
    |\hat{\psi}^{m,n} \rangle= C_m |\psi_{m,0}\rangle+ C_n |\psi_{n,0}\rangle \longleftrightarrow |\hat{\Psi}^{m,n} \rangle= \mathcal{N}_{m,0}C_m |\Psi_{m,0} \rangle +\mathcal{N}_{n,0}C_n |\Psi_{n,0} \rangle \,.
\end{align}
The Einstein's equation is linear in $O(G_N)$. Hence we find the perturbed metric for the superposed state $|\hat{\Psi}^{m,n} \rangle$ for a particular $m,n$. Finally, the complete perturbed metric due to the state \eqref{eq: primary state bulk} can be evaluated by summing properly over  $m$ and $n$. The details of the stress tensor components and the solution of the perturbed metric are provided in Appendix \ref{app: backreaction}. We keep in mind that at constant time the $rr$ and $r \varphi$ components of the metric perturbation are required to compute the perturbed area.

\subsection{Perturbed Area}
The unperturbed minimal area can be found from the geodesic connecting two end points of the interval in boundary CFT at a constant time slice \cite{Ryu:2006bv} 
\begin{align}\label{eq: area unperturbed}
    A_0[g]= 2\int_{r_{min}}^\infty d r \sqrt{\frac{1}{1+r^2}+r^2 \left(\frac{d \varphi}{ d r}\right)^2}\,.
\end{align}
Here $r_{min}$ is the radial distance of the turning point of the minimal curve (see Figure \ref{fig: Rindler}), which is related to the boundary interval $A_t$ of size $2 \pi x$ as follows:
\begin{align}\label{eq: pix and rm relation}
    \cot{\pi x} = r_{min}\,.
\end{align}
We want to compute the vacuum-subtracted entanglement entropy of the subsystem $A_t$ on any arbitrary time slice $t$ when the spacetime is backreacted with the perturbed metric. The expressions of the perturbed area are given in \eqref{eq: metric ansatz}. The perturbation breaks both the temporal and the angular symmetry of the spacetime. Hence, we need to calculate the area for the following two branches separately \cite{Chowdhury:2024fpd} (see Fig. \ref{fig: Rindler}).
\begin{subequations}\label{eq: slope branch}
\begin{align}
    & \text{Branch-I}:(\text{towards}\,\, r=r_{min}),\,\,\frac{d \varphi_I}{ d r}= \frac{r_{min}}{r \sqrt{(r^2-r_{min}^2)(1+r^2)}},\\
    & \text{Branch-II}:(\text{towards}\,\,r \rightarrow \infty),\,\,\frac{d \varphi_{II}}{ d r}=-\frac{r_{min}}{r \sqrt{(r^2-r_{min}^2)(1+r^2)}}\,.
\end{align}
\end{subequations}
As mentioned before, at constant time, only $rr$ and $r \varphi$ components of the perturbed metric are non-zero. For vacuum-subtracted entropy, we need to consider the change in the minimal area w.r.t. the unperturbed area \eqref{eq: area unperturbed}. In our case, this is given by
\begin{align}\label{eq: area formula}
    \delta A = & \int_{\gamma_A} \left[\frac{1}{1+r^2+\delta g_{rr}}+r^2 \left(\frac{d \varphi}{ d r}\right)^2+ 2\delta g_{r \varphi} \frac{d \varphi}{ d r}\right]^{\ha}- A_0[g]\\
    = & \notag \ha \left[ \int_{\infty}^{r_{min}} 2 \delta_I g_{r\varphi} \frac{r_{min}}{r^2}-\int^{\infty}_{r_{min}} 2 \delta_{II} g_{r\varphi} \frac{r_{min}}{r^2} -2 \int_{r_{min}}^\infty \delta g_{rr} \frac{\sqrt{r^2-r_{min}^2}}{r(1+r^2)^{3/2}}\right]\,.
    \end{align}
$\delta_I g_{r\varphi}$ and $\delta_{II} g_{r\varphi}$ denote the $r\varphi$ components of the metric in branches I and II respectively. The relative sign between the first two terms is due to the slope of the geodesic in two branches. The factor 2 in the third term comes from the angular symmetry of $\delta g_{rr}$. The $\varphi(r)$ in the two branches can be obtained integrating \eqref{eq: slope branch}.  The corresponding integration constants can be obtained demanding continuity of the geodesic in the two branches at $r=r_{min}$, and aligning the origin of the interval, i.e., the end point of branch-II with the origin of the $\varphi$ coordinate. Their expressions for the interval $(2 \pi a, 2 \pi a + 2 \pi x)$ are given by
\begin{align}\label{eq: varphi in branches}
    & \varphi_I = - 2\pi a + \arctan \frac{1}{r_{min}} -\arctan \left[ \frac{\sqrt{r^2-r_{min}^2}}{\sqrt{1+r^2} r_{min}}\right],\\
    & \notag \varphi_{II} =-2\pi a+ \arctan \frac{1}{r_{min}}  +\arctan \left[ \frac{\sqrt{r^2-r_{min}^2}}{\sqrt{1+r^2} r_{min}}\right].
\end{align}
We need the following functions in the computation of the perturbed minimal area.
\begin{align}\label{eq: cos sin in branch}
   & \notag \cos (\varphi_{I}+2\pi a)=-\frac{\sqrt{r^2-r_{min}^2}-\sqrt{r^2+1} r_{min}^2}{r \left(r_{min}^2+1\right)},\ \sin (\varphi_{I}+2\pi a)=\frac{r_{min} \left(\sqrt{r^2-r_{min}^2}+\sqrt{r^2+1}\right)}{r \left(r_{min}^2+1\right)}\,,\\
    & \notag \cos (\varphi_{II}+2\pi a)=\frac{\sqrt{r^2+1} r_{min}^2+\sqrt{r^2-r_{min}^2}}{r(1+ r_{min}^2)},\ \sin (\varphi_{II}+2\pi a)=-\frac{r_{min} \left(\sqrt{r^2-r_{min}^2}-\sqrt{r^2+1}\right)}{r \left(r_{min}^2+1\right)}\,.\\
\end{align}
Because of the short distance approximation, we can assume $\cos (\varphi+2\pi a)$ and $\sin (\varphi+2\pi a)$ as positive in both branches. Currently we will continue with $a=0$, at the end of this section we will generalize for arbitrary $a$.

Before proceeding, we would like to recall the entanglement entropy for primary excitation $|\Psi_{0,0} \rangle$ obtained in \cite{Belin:2018juv}. The back-reaction is time independent in this case, and the $rr$ and $tt$ components of the metric are perturbed. The perturbed area up to sub-leading order is given by
\begin{align}\label{eq: area primary}
    \delta A|_{\Psi_{0,0}}= 16 G_N h (1- r_{min} \cot^{-1} r_{min})-4 G_N (\pi x)^{4h} \frac{\Gamma \left(\frac{3}{2}\right) \Gamma\left(2h+1\right)}{\Gamma\left(2h+ \frac{3}{2}\right)}\,,
\end{align}
and the bulk entanglement entropy up to the sub leading order is given by
\begin{align}\label{eq: bulk EE primary}
    S_{bulk}(\Sigma_A)|_{\Psi_{0,0}}= (\pi x)^{4h} \frac{\Gamma \left(\frac{3}{2}\right) \Gamma\left(2h+1\right)}{\Gamma\left(2h+ \frac{3}{2}\right)}-(\pi x)^{8h} \frac{\Gamma \left(\frac{3}{2}\right) \Gamma\left(4h+1\right)}{\Gamma\left(4h+ \frac{3}{2}\right)}\,.
\end{align}
The sub leading order term in the perturbed area cancels with the leading order term in bulk entanglement entropy due to gravitational Gauss law \cite{Belin:2018juv}. We will note a similar pattern in our analysis.

\subsubsection*{Perturbed area for bulk dual state $|\hat{\Psi}\rangle$}

We start with the expression of the normalized bulk dual state
\begin{align} \label{eq: NH state}
    |\hat{\Psi}\rangle = \frac{1}{\sqrt{\hat{N}}}\sum_{q=0}^\infty \frac{(-i \tanh{\sigma})^q}{q!} \mathcal{N}_{q,0} a_{q,0}^\dagger |\Psi_{0,0} \rangle, 
\end{align}
where $\hat{N}$ is obtained as
\begin{align}
    \hat{N} =\left(1- \tanh^2 \sig \right)^{-2 h}\,.
\end{align}
We can write down the perturbed metric substituting  $C_m=\frac{(-i \tanh \sigma)^m}{m!}$ in \eqref{eq: metric ansatz}. During computing area in \eqref{eq: area formula}, we require the $rr$ and $r\varphi$ components of the perturbed metric, which we collect from Appendix \ref{app: backreaction}.
\begin{subequations}\label{eq: metric non-heating}
\begin{align}
& \delta g_{r r}^{(1)}(t,r,\varphi)\\
& \notag =  \frac{1}{\hat{N}}\sum_{m=0}^\infty |\mathcal{K}_m|^2 \left|\frac{(-i \tanh{\sigma})^{m}}{m!}\right|^2 d_m(r),\\
& \notag \delta g_{r \varphi}^{(1)}(t,r,\varphi)\\
& \notag = \frac{1}{\hat{N}}\sum_{m,n=0,m>n}^\infty \left[ R_{5 s}(r)+m n R_{5 p}(r)\right] (m-n) \mathcal{G}_{m,n}^{0}(\frac{3 \pi}{2}+t+\varphi)\\
& \notag = \frac{1}{\hat{N}}\sum_{m,n=0,m>n}^\infty \mathcal{K}_m \mathcal{K}_n \frac{(-i \tanh{\sigma})^{m+n}}{m! n!}\left[ R_{5 s}(r)+m n R_{5 p}(r)\right] (m-n) \mathcal{G}^0_{m,n}(t+\varphi),
\end{align}
where $\mathcal{K}_m$ is defined in \eqref{eq: Km def}, and also from the expression of $\mathcal{G}_{m,n}(x)$ in \eqref{eq: Gmn def} we can obtain
\begin{align} \label{eq: def G0}
    \mathcal{G}^0_{m,n}(x)&=\left[(-1)^{m}+(-1)^n\right]\sin [(m-n)x],~m-n=\text{even}\,,\\
    & \notag =-\left[(-1)^{m}-(-1)^n\right]i \cos [(m-n)x], ~ m-n=\text{odd}\,.
\end{align}
We will use the definition of $\mathcal{G}^0_{m,n}(x)$ while computing area in appendix \ref{App: details area}. We write down the metric components from \eqref{eq: metric solution} and \eqref{eq: metric large r}.
{\small
\begin{align}\label{eq: metric solution main text}
& d_m(r) = \tilde{D}_m-16 \pi  (2 h+m) r^{2 m} \, _2F_1\left(m,2 h+m;m+1;-r^2\right)+32 \pi  h r^{2 m}\left(r^2+1\right)^{-2 h-m+1},
\end{align}
\begin{align}
& \notag R_{5 s}(r)+m n R_{5 p}(r)\\
& \notag =\frac{1}{(m-n)^2\sqrt{r^2+1}}\bigg[(m-n)^2 (\tilde{c}_p^{m,n} m n+\tilde{c}_s^{m,n})- 8 \pi r^{m+n+1} \\
& \notag \times \bigg\lbrace 2h \left(r^2+1\right)^{\frac{1}{2} (-4 h-m-n+1)}-\frac{2
   (h (m+n+1)+m n){}_2F_1\left[\frac{(m+n+1)}{2} ,\frac{(4 h+m+n+1)}{2};\frac{(m+n+3)}{2};-r^2\right]}{m+n+1}\bigg\rbrace\bigg]\,.
\end{align}
}
\end{subequations}
These metric components are regular at large r for $h> \ha$, for certain choice of the integration constants. We discuss this in detail at the end of Appendix \ref{app: backreaction}. The details of the solution of Einstein's equation are provided in Appendix \ref{app: backreaction} and the associated ancillary notebook.
\normalsize
\paragraph*{Determination of $\tilde{D}_m$, $\tilde{c}_p^{m,n}$ and $\tilde{c}_s^{m,n}$:}~\\
In order to perform the sum over $m$ and $n$, we need to find the value of integration constants, $\tilde{D}_m$, $\tilde{c}_p^{m,n}$ and $\tilde{c}_s^{m,n}$. We use the Fefferman-Graham expansion of the metric to find the $tt$ component of the holographic stress tensor (see Appendix \ref{App: FG expansion}). This should match the stress tensor of the dual state in the boundary CFT. The boundary stress tensor for an arbitrary state $|\psi \rangle$ on a cylinder can be found as usual
\begin{align}
    \frac{\langle \psi| \mathcal{T}_{tt}|\psi \rangle}{\langle \psi | \psi \rangle}= \frac{1}{\langle \psi | \psi \rangle} \left[\langle \psi|\sum_{n=-\infty}^\infty \left( L_n e^{i n (t+\varphi)} +\bar{L}_n e^{-i n (t+\varphi)}\right)| \psi \rangle\right]\,.
\end{align}
We have computed this in the appendix \ref{App: boundary stress tensor}. We write down the integration constants found in \eqref{eq: metric large r} and \eqref{eq: integration const}.
\begin{align}\label{eq: integration const}
   & \tilde{D}_m=-8(2h+m) \frac{\mathcal{N}_{m,0}^2}{\mathcal{K}_m^2}+\frac{16 \pi  (2 h+m) \Gamma (2 h) \Gamma (m+1)}{\Gamma(2 h + m)},\\
   & \notag  (m n \tilde{c}^p_{m,n}+ \tilde{c}^s_{m,n} )\\
   & \notag = 4 (-1)^{m-n}\frac{[h (m-n+1)+n]}{(m-n)^2 \mathcal{K}_m \mathcal{K}_n} \frac{\Gamma (2 h+n) \Gamma(m+1)}{\Gamma (2 h)}- ( mn+ h(1+m+n))\frac{8 \pi  \Gamma (2 h) \Gamma \left(\frac{1}{2} (m+n+1)\right)}{(m-n)^2 \Gamma \left(\frac{1}{2} (4 h+m+n+1)\right)}.
\end{align}
Finally, we substitute these integration constants in \eqref{eq: metric solution main text} to obtain the metric perturbations.
\begin{align}\label{eq: metric non-heating colored}
& d_m(r)  = -8(2h+m) \frac{\mathcal{N}_{m,0}^2}{\mathcal{K}_m^2}\textcolor{red}{+32 \pi  h r^{2 m}\left(r^2+1\right)^{-2 h-m+1}}\nn
& \textcolor{blue}{+\frac{16 \pi  (2 h+m) \Gamma (2 h) \Gamma (m+1)}{\Gamma(2 h + m)}-16 \pi  (2 h+m) r^{2 m} \, _2F_1\left(m,2 h+m;m+1;-r^2\right)},\\
& \notag R_{5 s}(r)+m n R_{5 p}(r)\nn
& = 4 (-1)^{m-n}\frac{[h (m-n+1)+n]}{(m-n)^2 \mathcal{K}_m \mathcal{K}_n} \frac{\Gamma (2 h+n) \Gamma(m+1)}{\Gamma (2 h)}\frac{1}{\sqrt{r^2+1}} \textcolor{red}{-\frac{16\pi h}{(m-n)^2} r^{m+n+1}(r^2+1)^{\ha(-4h-m-n)}}\nn
& \textcolor{blue}{+\frac{r^{m+n+1}}{\sqrt{r^2+1}} \frac{16 \pi(h(m+n+1)+m n)}{(m+n+1)(m-n)^2} {}_2F_1\left[\frac{(m+n+1)}{2} ,\frac{(4 h+m+n+1)}{2};\frac{(m+n+3)}{2};-r^2\right]}\nn
& \textcolor{blue}{-\frac{m n+h(1+m+n)}{\sqrt{1+r^2}}\frac{8 \pi  \Gamma (2 h) \Gamma \left(\frac{1}{2} (m+n+1)\right)}{(m-n)^2 \Gamma \left(\frac{1}{2} (4 h+m+n+1)\right)}}.
\end{align}

We have decomposed the above metric perturbations into the three distinct kinds as follows:

1)\textit{Type-1 term}: Term marked in black, denoted by $\delta_1 g_{\mu\nu}$. The corresponding area is analytic in $r_{min}$ and can be found in the closed form expression, as we discuss in the next subsection. Area obtained from this type metric corresponds to the universal term $S_{uni}$ in the boundary CFT.

2) \textit{Type-2 term}: Term marked in red, denoted by $\delta_2 g_{\mu\nu}$ is an non-analytic function of $r_{min}$. The leading term of the corresponding area is $O(r_{min}^{-4h})$. This term cancels with the leading order term in the bulk entanglement entropy in \eqref{eq: bulk EE 1st}.

3)\textit{Type-3 term}: Term marked in blue, denoted by $\delta_3 g_{\mu\nu}$. The corresponding area is also non-analytic in $r_{min}$. However, we find that the leading order term in this part of area starts from $O(r_{min}^{-4h-2})$ (see details in the accompanying notebook). For non-integer $h$, we can ignore these terms in this paper.

We state the final expression of the area corresponding to type-1 and 2 terms as follows. The details are provided in Appendix \ref{App: details area} and the accompanying notebook.

\subsubsection*{Perturbed area: $rr$ component} We substitute $\delta g^{(1)}_{rr}(t,r,\varphi)$,\eqref{eq: metric non-heating} in the third term of \eqref{eq: area formula}. 
We find the type-1 and 2 area in terms of half of the boundary interval $\pi x$.
\begin{align}\label{eq: area rr NH type1}
    \delta_1 A_{rr}= \frac{16 G_N h \left(\pi x \cot \pi x-1\right)}{\tanh^2\sig-1}\,,
\end{align}
\begin{align}\label{eq: area rr NH type2}
\delta_2 A_{rr}= & -\frac{1}{\hat{N}}\sum_{m=0}^{\infty} \bigg[ |\mathcal{K}_m|^2 \left|\frac{(-i \tanh \sig)^{m}}{m!}\right|^2  \\
& \notag \times 4 G_N \pi ^{3/2} r_{min}^{-4 h} \frac{\Gamma (2 h+1)}{\Gamma (2 h+\frac{3}{2})} \, _2{F}_1\left(2 h,2 h+m+\frac{1}{2};2h+\frac{3}{2};-\frac{1}{r_m^2}\right)\bigg].
\end{align}
In $\delta_2 A_{rr}$, we keep only the leading order term in small $\pi x$ expansion which is given by
\begin{align}
    \delta_2 A_{rr}= G_N (\pi x)^{4 h} \left(-\frac{4 \pi ^{3/2} \Gamma (2h+1)}{\Gamma \left(2 h+\frac{3}{2}\right)}+O \left( (\pi x)^2\right)\right)\frac{1}{\hat{N}}\sum_{m=0}^{\infty} |\mathcal{K}_m|^2 \left|\frac{(-i \tanh{\sigma})^{m}}{m!}\right|^2 .
\end{align}
Note that $\delta_2 A_{rr}$ matches the corresponding area found in eq.(4.11) of \cite{Belin:2018juv} for $m=0$, and with the eq. (D.25), (D.34) in \cite{Chowdhury:2024fpd} for $m=1$ and $2$, after adjusting the normalization factor suitably.
 
\subsubsection*{Perturbed area: $r\varphi$ component} 

Similarly, substituting $\delta g^{(1)}_{r \varphi}$, \eqref{eq: metric non-heating} in the first and second term of \eqref{eq: area formula} we can compute the area. The perturbed metric has dependence on $t+\varphi$, so we need to use \eqref{eq: cos sin in branch} to write the integral in \eqref{eq: area formula} in terms of $r$. In addition, $\mathcal{G}^0_{m,n}(t+\varphi)$ has a different form for odd or even $(m-n)$. Hence, we compute the area for several cases to observe the pattern. The area has been described in Appendix \ref{App: details area}. We provide the final expression below.
\begin{align}\label{eq: area r varphi NH type1 main}
    \delta_1 A_{r\varphi} & = -\frac{8 G_N h \tanh{\sigma} (\sin 2 \pi  x-2 \pi  x) \csc \pi  x \sin (t+\pi  x)}{\tanh^2{\sigma}-1}\\
    & \notag +\frac{4 G_N h}{\tanh^2{\sigma}-1} \bigg[ 4 \tanh^2{\sigma} + 4 \tanh{\sigma} \cos \pi  x \sin (t+\pi  x) \\
    & \notag+\left \lbrace 2 i \tanh{\sigma}\cos t + i \cot \pi  x \left(2 \tanh{\sigma} \sin t+ \tanh^2{\sigma}+1\right)\right \rbrace\left( L(t+2 \pi x)-L(t) \right)\bigg],
    \end{align}
where we have defined 
\begin{align}\label{eq: def L(t)}
L(y)=\log \frac{\left(1-i \tanh{\sigma} e^{i y} \right)}{\left(1+i \tanh{\sigma} e^{-i y} \right)}\,.
\end{align}
We combine \eqref{eq: area rr NH type1} and \eqref{eq: area r varphi NH type1 main} to obtain the closed form expression of type-1 perturbed area.

Next, we state the leading term of the corresponding area for the type 2 metric found in \eqref{eq: area r varphi type 2}:
\begin{align}\label{eq: area r varphi type 2 main}
   \delta_2 A_{r\varphi}^{NH} & =  -\frac{1}{\hat{N}_{NH}}G_N (\pi x)^{4 h}\frac{8 h \pi ^{3/2} \Gamma (2
   h)}{\Gamma \left(2 h+\frac{3}{2}\right)}  \\
   & \notag \bigg\lbrace \sum_{m,n=0,m-n=odd}^\infty  -i \frac{(-i \tanh{\sigma})^{m+n}}{m! n!} [(-1)^m-(-1)^n] \sin[(m-n)t] \mathcal{K}_{m}\mathcal{K}_{n}\\
   & \notag + \sum_{m,n=0,m-n=even}^\infty\frac{(-i \tanh{\sigma})^{m+n}}{m! n!} [(-1)^m+(-1)^n] \cos[(m-n)t]\mathcal{K}_{m}\mathcal{K}_{n}\bigg \rbrace.
\end{align}
We do not attempt to perform the sum over $m,n$, as we will see in the next subsection that at $O((\pi x)^{4h})$ this cancels with  the bulk entanglement entropy.

\subsection{Bulk entanglement entropy}
We would like to compute the bulk entanglement entropy of an interval $(2 \pi a, 2\pi a+2 \pi x)$ at an arbitrary time slice $t=t'$. We need to choose an AdS-Rindler co-ordinate transformation \cite{Casini:2011kv,Belin:2018juv}, such that the minimal surface of an boundary interval $(2 \pi a, 2\pi a+ 2\pi x)$ at $t=t'$ is mapped to the Rindler horizon. We start with the following Rindler transformation.
\begin{align} \label{coordchange}
 & r=\sqrt{\rho^2 \sinh^2 x + \left( \sqrt{\rho^2-1} \cosh \eta \cosh \tau + \rho \cosh x \sinh \eta\right)^2} \, , \\ 
&t-t'= \arctan\left( \frac{\sinh \tau \sqrt{\rho^2-1}}{\rho \cosh x \cosh \eta +\sqrt{\rho^2-1} \cosh\tau \sinh \eta}\right)  \nonumber \, , \\ 
\nonumber &\varphi -\frac{\hat{\theta}}{2}+2\pi a=\arctan\left( \frac{\rho \sinh x}{\sqrt{\rho^2-1} \cosh \tau \cosh \eta +\rho \cosh x \sinh \eta}\right) \,,
\end{align} 
where
\[\hat{\theta}=2 \pi x\]
denotes the boundary interval, and the parameter $\eta$ is chosen such that
\begin{equation} \label{relcosh}
\cosh \eta = \frac{1}{\sin \frac{\hat{\theta}}{2} } = \sqrt{ r_{min}^2 +1}.
\end{equation}
The metric in AdS-Rindler coordinate is given by
\begin{equation} \label{rinbtz}
ds^2 = -(\rho^2 -1) d\tau^2 + \frac{d\rho^2}{ \rho^2 -1} + \rho^2 dx^2, \qquad x \in \mathbb{R},\, \tau \sim \tau + 2\pi i\,.
\end{equation}
Using the identification \eqref{relcosh}, we can show that the Rindler horizon is a surface in the co-ordinates $(r,\varphi)$ governed by the equation below.
\begin{equation}
r^2 = \sinh^2 x + \cosh^2 x \sinh^2 \eta, \qquad \tan ( \varphi -\frac{\hat{\theta}}{2}+ 2 \pi a )=  \frac{\sinh x}{ \cosh x \sinh \eta}.
\end{equation}
Using the above equations and replacing $\varphi(r)$, we can write the following for the two branches of the minimal surface or horizon.
\begin{align}
   & \tan (\varphi_{I}- \frac{\hat{\theta}}{2}+2\pi a)=\frac{\sqrt{r^2-r_{min}^2}}{r_{min} \sqrt{1+r^2}}
   & \tan (\varphi_{II}- \frac{\hat{\theta}}{2}+2\pi a)=- \frac{\sqrt{r^2-r_{min}^2}}{r_{min} \sqrt{1+r^2}}
\end{align}
We can check that the above equations are consistent with \eqref{eq: cos sin in branch}. Currently, we assume $a=0$. The minimal surface splits the bulk in the two wedges (see Figure \ref{fig: Rindler}). The wedge corresponding to the boundary subsystem is denoted as the right wedge, while the wedge corresponding to the complimentary subsystem in the boundary is denoted as the left wedge. Hence, the field $\phi(\tau,\rho,x)$ can be expanded in the left and right Rindler modes $g(\om,k,I)$
\begin{align}
  \phi(\tau,\rho,x)= \sum_{I \in L,R}\int_{0}^\infty \frac{d \om}{2 \pi}  \int_{-\infty}^\infty \frac{d k}{2 \pi} \left[ g_{\om,k,I} e^{i k x} e^{-i \om t} b_{\om,k,I}+ g^*_{\om,k,I} e^{-i k x}e^{i \om t} b^\dagger_{\om,k,I} \right]\,.
\end{align}
The Rindler creation(annihilation) operators follow the standard commutation relation. The expressions of the left and right Rindler modes are given by \cite{Belin:2018juv}
\begin{align}\label{eq: Rindler mode}
g_{\om,k,I}= N_{\om,k} e^{i k x} \rho^{-2h}(1-\frac{1}{\rho^2})^{-\frac{i\om}{2}} {}_2F_1 \left[ h-\frac{i(\om+k)}{2}, h- \frac{i(\om-k)}{2};2h; \frac{1}{\rho^2}\right]\,.
\end{align}
The normalization constant $N_{\om,k}$ is given by
\begin{align}\label{eq: Rindler normalization}
    N_{\om,k}= \frac{1}{\sqrt{2 \om}}\frac{\left| \Gamma \left( h+ \frac{i(k+\om}{2}\right) \Gamma \left( h+ \frac{i(k-\om)}{2}\right) \right|}{\Gamma(2h) |\Gamma(i \om) | }\,.
\end{align}
The creation(annihilation) operators in the two coordinates are related by the \bbgv coefficients.
\begin{align}
    a_{m,n}= \sum_{I,\om,k} \left( \alpha_{m,n;\om,k,I} b_{\om,k,I} + \beta^*_{m,n;\om,k,I} b^\dagger_{\om,k,I} \right)\,.
\end{align}
It should be noted that $\om$ and $k$ being continuous, the summation symbol used above acts as a representation of the integration over $\om$ and $k$. In the Rindler coordinate, the global vacuum is a thermofield double state when written using the Rindler energy eigenstates. Using this fact, one can find the relation between the creation(annihilation) operators in the two wedges, when acting on the global vacuum. We also get an orthogonality constraint of the \bbgv coefficients in terms of right wedge quantities.
\begin{align}\label{eq: bbgv orthogonality}
    \sum_{\om,k} \left[ |\alpha_{m,n;\om,k,R}|^2 (1-e^{-2 \pi \om})-|\beta_{m,n;\om,k,R}|^2 (1-e^{2 \pi \om}) \right]=1\,.
\end{align}
Most importantly, the excited states in the global coordinate can be written in terms of a specific superposition of excited states, solely in terms of the right wedge quantities \cite{Belin:2018juv, Chowdhury:2024fpd}.
\begin{align}
    |\Psi_{m,n} \rangle  & \equiv a_{m,n}^\dagger |0 \rangle\\
    & \notag = \sum_{\om,k} \left[ (1- e^{-2 \pi \om}) \alpha^*_{m,n;\om,k,R} b^\dagger_{m,n;\om,k,R} + (1- e^{2 \pi \om}) \beta_{m,n;\om,k,R} b^\dagger_{m,n;\om,k,R} \right] |0 \rangle\,.
\end{align}
This is the key relation to find the bulk entanglement entropy as it trivializes the computation of reduced density matrix. Henceforth, we will drop the index $R$ as we will only use the right wedge quantities.  We consider the change in the $q$-th Renyi entropy w.r.t the vacuum value.
\begin{align}
 \delta S_q^{bulk}= \frac{1}{1-q} \log \frac{ Tr \rho_1^q}{Tr \rho_0^q }   \,,
\end{align}
where $\rho_1$ is the reduced density matrix of the excited state and $\rho_0$ is that of the vacuum. In the end, we take $q \rightarrow 1$ analytically to find the Von-Neuman entropy. We chose the following perturbation parameter for the computation of the bulk entanglement entropy
\begin{align}\label{eq: perturbation bulk EE}
    \delta \rho= \rho_1-\rho.
\end{align}
 This is because $\delta \rho \rightarrow 0$ in the short distance approximation. We compute the entanglement entropy up to the second order in $\delta \rho$. It is to be noted that simply performing the small $\pi x$ or large $\eta$ expansion of $\rho_1$ is not the correct way to take the short distance approximation. The \bbgv coefficients maintain the orthogonality constraint \eqref{eq: bbgv orthogonality} among themselves, which implies that the \bbgv coefficients might not always be small in the large $\eta$ limit  when integrated over $\om$ and $k$ \cite{Belin:2018juv}. 

In the following, we proceed to find the bulk entanglement entropy of $|\hat{\Psi} \rangle$ up to the second order in the short distance approximation. The bulk entanglement entropy is proportional to that of the primary state $|\Psi_{0,0} \rangle$ in this limit,, which is expected from the Virasoro symmetry\cite{Chowdhury:2021qja,Chowdhury:2024fpd}. 

\subsection*{Short distance approximation: first order}
Our bulk state has the following form
\begin{eqnarray} \label{exstateright}
 |\Psi\rangle &=& \sum_{m, n} B_{m, n} a^\dagger_{m,n} |\Psi_{0,0}\rangle, 
 \end{eqnarray}
 The coefficients $B_{m,n}$ are such that the state is normalized to unity, i.e.,
\begin{eqnarray}
 \sum_{mn} |B_{m,n}|^2 =1.
 \end{eqnarray}
Then the first order correction to the bulk entanglement entropy can be obtained as \cite{Chowdhury:2024fpd}
  \begin{eqnarray} \label{1stsbulk}
    S_{bulk}^{(1)}  (\Sigma_A)  = 2\pi \sum_{\omega, k} \omega \Big( | B\cdot\alpha^* |^2 
    + |B\cdot \beta|^2  \Big),
  \end{eqnarray}
where we have defined
  \begin{equation}
  B\cdot \alpha^*  = \sum_{m, n}B_{m, n}  \alpha^*_{m, n; \omega, k}, \qquad
  B\cdot \beta =\sum_{m, n} B_{m, n} \beta_{m, n; \omega, k}.
  \end{equation}
From \eqref{eq: NH state}, we note the non-zero superposition coefficients $B_{m,n}$.
\begin{equation}
    B_{n,0}= \frac{1}{\sqrt{\hat{N}}}\frac{(-i \tanh{\sigma})^n}{n!}\mathcal{N}_{n,0}\,.
\end{equation}
 In Appendix \ref{app: bbgv coefficients}, we have noted the scaling relation \eqref{eq:bbgv scaling} of the \bbgv coefficients of the descendants w.r.t the primary excitation in the short distance approximation and $|\alpha_{0,0;\om,k}|=|\beta_{0,0;\om,k}|$. Also, at the leading order, in the short distance approximation, we obtain
\begin{align}\label{eq: bbgv approx}
    \lim_{\eta \rightarrow \infty} B\cdot\alpha^*  = \lim_{\eta \rightarrow \infty} B\cdot\beta= \frac{1}{\sqrt{\hat{N}}} \lim_{\eta \rightarrow \infty}\sum_{m=0}^\infty \frac{(-i \tanh{\sigma})^m}{m!} e^{- i m (t'+ \pi x)}  \alpha_{0,0;\om,k} \frac{\mathcal{K}_m}{N_{0,0}},
\end{align}
where we have taken the origin of the interval at $a=0$ and $\mathcal{K}_m$ is defined in \eqref{eq: Km def}. From \eqref{eq: normalization const bulk}, we note $N_{0,0}= \frac{1}{\sqrt{2\pi}}$. Hence, the first order bulk entanglement entropy of the state $|\hat{\Psi} \rangle$ at time $t$ becomes
\begin{align}\label{eq: bulk EE 1st}
   & S_{bulk}^{(1)}(\Sigma_{A_t})|_{\hat{\Psi}}\\
   & \notag =  S_{bulk}^{(1)}(\Sigma_{A_t})|_{\Psi_{0,0}} \times \frac{2 \pi}{\hat{N}} \left|\sum_{m=0}^\infty \frac{(-i \tanh{\sigma})^m}{m!} e^{- i m (t+\pi x)} \mathcal{K}_m \right|^2\\
   & \notag =\frac{\pi^{\frac{3}{2}} \Gamma(2h+1)}{\Gamma(2h+ \frac{3}{2})} \frac{(\pi x)^{4h}}{\hat{N}}\bigg\lbrace \sum_{m=0}^\infty \left|\frac{(-i \tanh{\sigma})^m}{m!}\right|^2 |\mathcal{K}_m|^2 \\
   & \notag -i \sum_{m,n=0,m-n=odd}^\infty \frac{(-i \tanh{\sigma})^{m+n}}{m! n!} [(-1)^m-(-1)^n] \sin[(m-n)(t+\pi x)] \mathcal{K}_m \mathcal{K}_n\\
   & \notag + \sum_{m,n=0,m-n=even}^\infty\frac{(-i \tanh{\sigma})^{m+n}}{m! n!} [(-1)^m+(-1)^n] \cos[(m-n)(t+\pi x)] \mathcal{K}_m \mathcal{K}_n\bigg \rbrace \,.
\end{align}
\emph{ At $O((\pi x)^{4h})$, the entropy cancels with the perturbed area due to the type-2 metric as can be seen from \eqref{eq: area rr NH type2} and \eqref{eq: area r varphi type 2 main}}.

\subsection*{Short distance approximation: second order}

The expression of the second order correction to the bulk entanglement entropy can be obtained as (see eq.(3.133) in \cite{Chowdhury:2024fpd})
\begin{eqnarray} \nonumber
&&S^{(2)}_{bulk}(\Sigma_{A_t})=
 \frac{1}{2} \sum_{\stackrel{\omega_1, \omega_2}{k_1, k_2} }  \Bigg[
2\pi  (\omega_1 - \omega_2) \frac{ ( 1-e^{-2\pi \omega_1} ) (1- e^{2\pi \omega_2} ) }{ 1- e^{2\pi ( \omega_2 - \omega_1)}}
  \Big|( B\cdot \alpha_1^* )(B^* \cdot \alpha_2) + (B^* \cdot \beta_1^*)( B\cdot \beta_2) \Big|^2
  \\ \nonumber
 && +2\pi ( \omega_1 + \omega_2) \frac{ (1-e^{2\pi \omega_1})  (1-e^{2\pi \omega_2})}{ 1- e^{ 2\pi (\omega_1 + \omega_2)}}
 2 \Big\{ | B\cdot \alpha_1^*|^2 |B^*\cdot \beta_2|^2 + (B\cdot \alpha_1^*) (B^* \cdot \beta_2^*) (B^* \cdot \alpha_2) ( B \cdot \beta_1) \Big\}
 \Bigg],
 \\  
\end{eqnarray}
where we have denoted $\alpha(\beta)_i= \alpha(\beta)_{m,n; \om_i,k_i}$. We follow the previous subsection to obtain the second order bulk entanglement entropy in short distance approximation of $|\hat{\Psi}\rangle$ at time $t$.
\begin{align} \label{eq: bulk EE 2nd}
    S^{(2)}_{bulk}(\Sigma_{A_t})|_{\hat{\Psi}} &= S^{(2)}_{bulk}(\Sigma_{A_t})|_{\Psi_{0,0}} \times \frac{(2 \pi)^2}{\hat{N}^2}\left| \sum_{m=0}^\infty \frac{(-i \tanh{\sigma})^m}{m!} e^{- i m t} \frac{\mathcal{K}_m}{N_{0,0}} \right|^4\\ 
    & \notag = S^{(2)}_{bulk}(\Sigma_{A_t})|_{\Psi_{0,0}} \left( \cosh 2 \sig + \sinh 2 \sig \sin (t+\pi x) \right)^{-4 h},
\end{align}
Note the scaling w.r.t the bulk entanglement entropy of the primary excitation.

The bulk entanglement entropy of the primary excitation can be calculated as follows \cite{Colin-Ellerin:2024npf, Bhat:2025iqb} (note that the integration of $\omega$ is from 0 to $\infty$) . We use \cite{NIST:DLMF} to compute the integration over $k_1$ and $k_2$.
\begin{eqnarray} \nonumber
S^{(2)}_{bulk}(\Sigma_{A_t}) && =
4 \pi \int \frac{d \om_1}{2 \pi} \frac{d \om_2}{2 \pi} \frac{d k_1}{2 \pi} \frac{d k_2}{2 \pi}   \Bigg[ (\omega_1 - \omega_2) \frac{ ( 1-e^{-2\pi \omega_1} ) (1- e^{2\pi \omega_2} ) }{ 1- e^{2\pi ( \omega_2 - \omega_1)}}
  |\alpha_{0,0;\om_1,k_1}|^2 |\alpha_{0,0;\om_2,k_2}|^2 \\ \nonumber
&& + ( \omega_1 + \omega_2) \frac{ (1-e^{2\pi \omega_1})  (1-e^{2\pi \omega_2})}{ 1- e^{ 2\pi (\omega_1 + \omega_2) } } |\alpha_{0,0;\om_1,k_1}|^2 |\alpha_{0,0;\om_2,k_2}|^2
 \Bigg].\\ \nonumber
&& = \frac{2^{8h} (\cosh \eta)^{-8h}}{4 \pi^2} \int \frac{d \om_1}{2 \pi} \frac{d \om_2}{2 \pi} \om_1 \om_2  |\Gamma( i \om_1)|^2 |\Gamma(i \om_2)|^2 \frac{|\Gamma(2h+i \om_1)|^2 |\Gamma(2h+i \om_2)|^2}{(\Gamma(4h))^2}\\ \nonumber
&& \times \Bigg[ (\omega_1 - \omega_2) \frac{ ( 1-e^{-2\pi \omega_1} ) (1- e^{2\pi \omega_2} ) }{ 1- e^{2\pi ( \omega_2 - \omega_1)}} + ( \omega_1 + \omega_2) \frac{ (1-e^{2\pi \omega_1})  (1-e^{2\pi \omega_2})}{ 1- e^{ 2\pi (\omega_1 + \omega_2) } }\Bigg]\,.
\end{eqnarray}
We further use the reflection formula of the Gamma function and \cite{gradshteyn2007} to finally obtain
\begin{align}
\nonumber S^{(2)}_{bulk}(\Sigma_{A_t}) &=  -\frac{2^{8h} (\cosh \eta)^{-8h} }{ \pi} \frac{\sqrt{\pi}}{2} \frac{\Gamma(2h) \Gamma(2h+1) \Gamma(2h+ \ha) \Gamma(2h+ \frac{3}{2}) \Gamma(4h+1)}{\Gamma(4h) \Gamma(4h+\frac{3}{2}) \Gamma(4h+2)}\\
& = -(\pi x)^{8h} \frac{\Gamma(\frac{3}{2})\Gamma(4h+1)}{\Gamma(4h+\frac{3}{2})}.
\end{align}

\subsection*{Holographic entanglement entropy} We combine the perturbed area $\delta A_{rr}$ in \eqref{eq: area rr NH type1}, $\delta A_{r\varphi}$ in \eqref{eq: area r varphi NH type1 main} and the bulk entanglement entropy in \eqref{eq: bulk EE 2nd} to write the full entanglement entropy of the bulk state $|\hat{\Psi} \rangle$ for the boundary interval $A_t=(0,2 \pi x)$, at time $t$, and up to second order in the short distance approximation.
\begin{align}\label{eq: final holo ee}
    S_{EE} &= \frac{\delta A_{rr}}{4 G_N}+ \frac{\delta A_{r\varphi}}{4 G_N}+ S^{(2)}_{bulk} (\Sigma_{A_t})+ \cdots  \\
    & \notag =\frac{4 h \left(\pi x \cot \pi x-1\right)}{\tanh^2 \sig-1} -\frac{2 h \tanh \sig (\sin 2 \pi  x-2 \pi  x) \csc \pi  x \sin (t+\pi  x)}{\tanh^2 \sig-1}\\
    & \notag +\frac{ h}{\tanh^2 \sig-1} \bigg[ 4 \tanh^2 \sig + 4 \tanh \sig \cos \pi  x \sin (t+\pi  x) \\
    & \notag +\left \lbrace 2 i \tanh \sig \cos t +i \cot \pi  x \left(2 \tanh \sig \sin t+\tanh^2 \sig+1\right) \right \rbrace\left( L(t+2 \pi x)-L(t) \right)\bigg]\\
    & \notag -(\pi x)^{8h} \frac{\Gamma(\frac{3}{2})\Gamma(4h+1)}{\Gamma(4h+\frac{3}{2})} \left( \cosh 2 \sig + \sinh 2 \sig \sin (t+\pi x) \right)^{-4 h}+ \cdots ,
\end{align}
where $\sigma= n \beta T$, and $L(t)$ is defined in \eqref{eq: def L(t)}.
\begin{align*}
L(y)=\log \frac{\left(1-i \tanh{\sigma} e^{i y} \right)}{\left(1+i \tanh{\sigma} e^{-i y} \right)}\,.
\end{align*}
One can check the entanglement entropy is real. At the sub-leading order in $\pi x$, this matches the entanglement entropy of the dual state $|\hat{\psi}\rangle$ in \eqref{eq: prepared state cft normal order main} in the boundary CFT, which has been evaluated in \eqref{eqn:cftuni} and \eqref{eqn:sdyn2}, hence, confirming the FLM conjecture. The type-1 term or the universal contribution of the entropy also matches with previous literature \cite{Caputa:2022zsr}, where the holographic entanglement entropy for a coherent state was computed in $h\sim c$ limit using the Banados geometry. In Appendix \ref{appendix I}, we have shown this in detail.

As a further check of our result, we expand the holographic entanglement entropy to the linear order in $ \sig= n\beta T$, which is equivalent to finding the entanglement entropy for the state $C_0 |\Psi_{0,0} \rangle + C_1 |\Psi_{1,0} \rangle$.
\begin{align}\label{eqn:holo_ee}
    S_{EE} & = 4 h (1-\pi  x \cot \pi  x) + 2 h n \beta T (\sin 2 \pi  x-2 \pi  x) \csc \pi  x \sin (t+\pi  x)\\
    & \notag -(\pi x)^{8h} \frac{\Gamma(\frac{3}{2})\Gamma(4h+1)}{\Gamma(4h+\frac{3}{2})}\left( 1-8 h n \beta T \sin (t+\pi x) \right). 
\end{align}
At $t=0$, and in the sub-leading order in $\pi x$, this matches the equivalent value of the entanglement entropy obtained in eq.(3.75) and (3.138) in \cite{Chowdhury:2024fpd}.

\begin{figure}[!h]
   \centering        
   \includegraphics[width=0.48\textwidth]{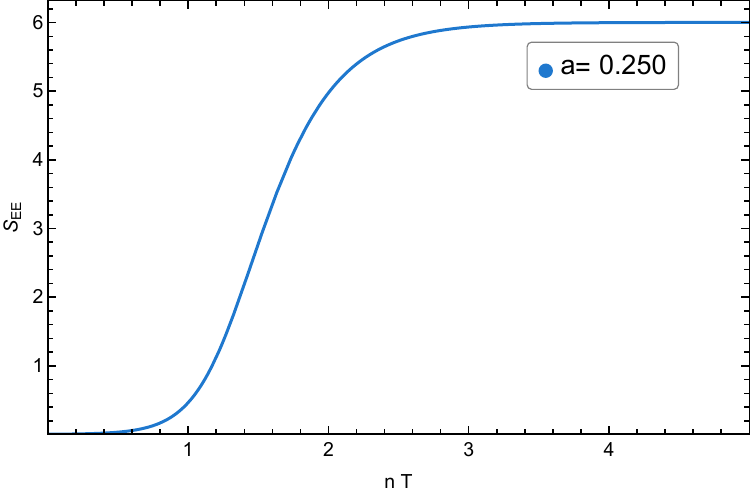} 
   \includegraphics[width=0.51\textwidth]{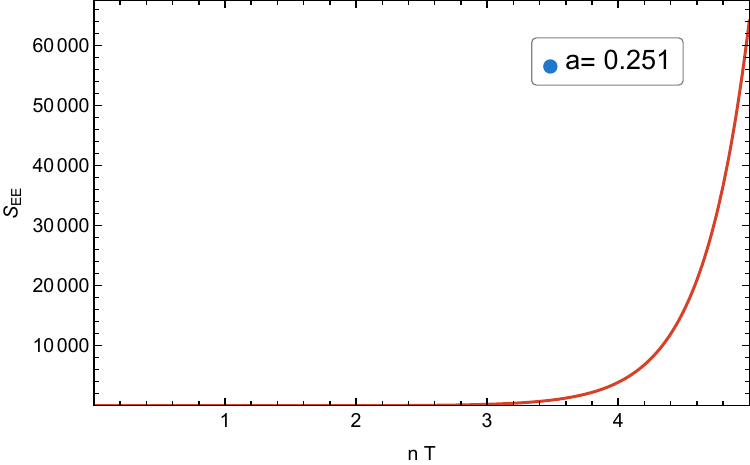}\\
   \hspace{0.5 in}
   \includegraphics[width=0.45\textwidth]{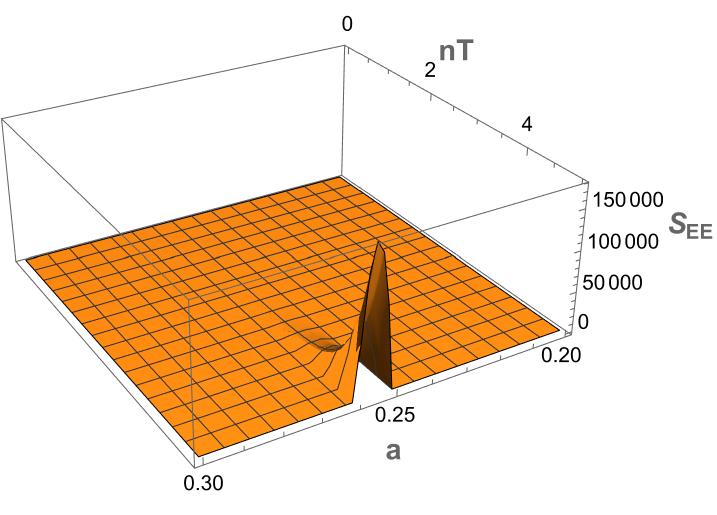}
   \includegraphics[width=0.45\textwidth]{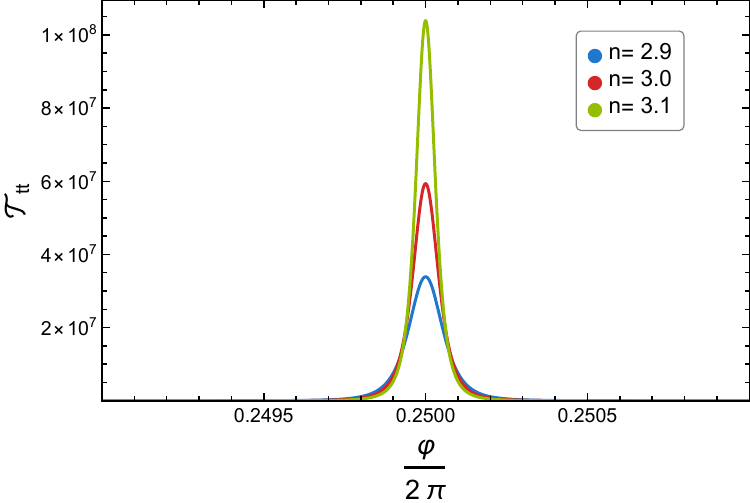}
   \caption{{\it Top: }The Leading order behavior of $S_{EE}$ in \eqref{eq: final holo ee} in the small $x$ limit, as a function of $nT$ for $ h=3, t=0, \beta=1.4, \alpha=0$ and $x=0.01$. We can see changing the origin of interval $a$ from $0.25$ to $0.251$ leads to sharp increase in the entanglement entropy, consistent with the peak in the boundary stress tensor $\mathcal{T}_{tt}$ . {\it Bottom left:} $S_{EE}$ as a function of $nT$ and $a$ for the same parameters in the top plot. {\it Bottom right:} Boundary stress tensor $\mathcal{T}_{tt}$ found in \eqref{eq: boundary stress tt full} as a function of $\varphi$ at $t=0, T=1, h=3, \gamma=1.4$ and $x=0.01$.  Due to the holomorphic drive in the bulk, we have only one peak where the energy and entropy diverges.}
   \label{fig: bulkplot}
\end{figure}

\paragraph{\underline{Interval $A_t=(2 \pi a, 2 \pi a+ 2 \pi x)$}:}~\\ We can easily generalize the above expression of $S_{EE}$ to an interval with arbitrary origin $a$. We note the expressions of $\varphi(r)$ \eqref{eq: varphi in branches} and the AdS-Rindler coordinate transformation in \eqref{coordchange} for non-zero $a$. From the presence of $\cos (m-n) (t+ \varphi)$ and $\sin(m-n)(t+\varphi)$ in the metric $\delta g_{r\varphi}$ in \eqref{eq: metric non-heating}, we can see that changing the interval from $(0,2\pi x)$ to $(2 \pi a, 2 \pi a+ 2 \pi x)$ is equivalent to changing the time coordinate $t \rightarrow \tilde{t}=t + 2\pi a$ in the metric. The final expression of the area at $\tilde{t}=0$ can be obtained by substituting $t \rightarrow - 2\pi a$ into the area term of the holographic entanglement entropy \eqref{eq: final holo ee}. The bulk entanglement entropy is already generalized for arbitrary $a$ in \eqref{eq:bbgv scaling}. However, because of the short distance approximation, we cannot take range of $a$ more than $0.5$, while plotting $S_{EE}$, as the value of $\cos \varphi$ and $\sin \varphi$ in \eqref{eq: cos sin in branch} changes sign in different quadrants of the circle in the boundary CFT.

\paragraph{\underline{Signature of Heating phase}:}~\\ We set out to find the signatures of dynamical phases with the expression of entanglement entropy. We have assumed $\alpha=0$ (see below for non-zero $\alpha$). The order parameter $\delta$ in \eqref{eqn:3phases} indicates that this corresponds to the heating phase. Consistently, we note that the saturating behavior of $S_{EE}$ at $a=0.25$ and the divergent behavior at $a=0.251$, both with interval size $x=0.01$ (see fig.\ref{fig: bulkplot}). Hence we conclude that the peak in the entanglement entropy lies around $a=0.25$, which is consistent with the peak in the boundary CFT stress tensor $\mathcal{T}_{tt}$.  Note that unlike in the CFT (fig.\ref{fig:ee_plot3d}), we have only one peak due to suppressed anti-holomorphic drive in the bulk. However, our result is not valid for large $nT \beta$, as we have computed the  backreaction at $G_N \rightarrow 0$ and also computed the bulk entanglement entropy perturbatively (see \eqref{eq: perturbation bulk EE}).

\paragraph{\underline{Inclusion of anti-holomorphic sector}:}~\\
The state in this case will be
\begin{align}\label{eq: prepared state cft normal order}
    |\hat{\psi}\rangle & =  e^{A_+^R\bar{L}_{-1}} e^{\bar{L}_0\log[A_0^L]} e^{A_-^R \bar{L}_1} e^{A_+^R L_{-1}} e^{L_0\log[A_0^L]} e^{A_-^R L_1}|h,h\rangle\\
    & \notag =  e^{2 h\log[A_0^R]} \bigg \lbrace \sum_{q, \bar{q}=0 {\substack{q \neq \bar{q}}}}^\infty \frac{(A_+^R)^q}{q!} \frac{(A_+^R)^{\bar{q}}}{\bar{q}!} \bar{L}_{-1}^{\bar{q}} L_{-1}^q |h,h \rangle +\sum_{p=0}^\infty \frac{(A_+^R)^p}{p!} \bar{L}_{-1}^{p} L_{-1}^p |h,h \rangle \bigg \rbrace
\end{align}
The first sum above will be the same as the holomorphic drive except for the fact that the angular momentum quantum number will also take values from $-\infty$ to $\infty$. The negative integers of angular momenta will correspond to the anti-chiral peak in the entanglement entropy. The second sum involves the states with zero angular momenta. It can be shown that these states will not break the angular symmetry of the bulk spacetime \cite{Chowdhury:2024fpd}, hence should remain as an additive constant.

\paragraph{\underline{Arbitrary $\alpha$}:}~\\ Finally, we can generalize our result for arbitrary values of the drive parameters using the general expression of $A_+^R$ as obtained in \eqref{eqn:normalorder} to construct our state \eqref{eq: primary state bulk}. The corresponding entanglement entropy $S_{EE}$ can be evaluated by replacing $ \tanh \sigma \rightarrow \frac{ n T\beta \frac{\sinh \sig}{\sig}}{\cosh \sig+i n T\alpha \frac{\sinh \sig}{2\sig}}$ in \eqref{eq: NH state}. The computation of perturbed area is straightforward but more involved in that case, hence we do not pursue it in this work.

\section{Discussion}\label{discussion}

In this paper, we compute the time evolution of the semiclassical entanglement entropy of a state prepared via a  $sl(2,\mathbb{R})$ valued drive in the small interval limit. On the CFT side, this computation is done for three different cases, corresponding to the drive being in the heating phase, non-heating phase and phase boundary. The effect of the drive can be captured by a local operator insertion at a conformally transformed point, which depends on the drive parameters. We note the divergent behavior of the entanglement entropy in the heating phase, and an oscillatory growth in the heating phase. The divergent behavior of the heating phase occurs only when our interval includes either of the energy peaks.

Next, we consider the bulk dual of this prepared state, for a specific case $(\alpha'=\beta'=\gamma' =0,\; \textrm{and}\; \alpha=0, \; \beta =\gamma)$, (which corresponds to a holomorphic drive) in the $AdS_3$. This is a superposed state of primary and all holomorphic global descendants with non-zero angular momenta, with the superposition coefficients containing the drive parameters and the stroboscopic time. We compute the time evolution of the perturbed minimal area, and the bulk entanglement entropy at bulk time $t$. The holographic entanglement entropy, thus obtained, matches the entanglement entropy in the boundary CFT. This leads to a nontrivial check of the Faulkner-Lewkowycz-Maldacena(FLM) formula. When we put $t=0$ in this expression, and vary the entanglement entropy w.r.t drive parameter $nT$, we observe the signature of the heating phase. Because of the holomorphic drive, we observe only one peak of entanglement entropy here.

Our results in the bulk are perturbative in nature where we need to keep the combination $n T \beta$ finite, hence we cannot increase both of the stroboscopic time and drive parameters to an arbitrary large value. Also, we did not compute the $O(c^1)$ entanglement entropy in the bulk. In crude way, it seems the area of the classical Ryu-Takayanagi surface will not have the drive parameter or stroboscopic time dependence as we are computing the area in the global coordinate. We probably need to use the non-uniform holographic cutoff in order to derive the $O(c^1)$ entanglement entropy (see \cite{Jiang:2024hgt} for details).

 The growth of the entanglement entropy we found here is a bit different from that in the existing literature, since we are considering $O(c^0)$ term with finite values of the drive parameters and stroboscopic time. But, it is expected that the semiclassical results will reflect the classical result, because the application of drive mixes the primary state with the descendants in such a way that it resembles a coherent state (at least for $\alpha=0$ case). The universal or type-1 term in the entanglement entropy alone shows the signature of the heating phase, hence our conclusion should be valid for non-holographic CFTs too. One can check whether these behaviors for the different phases are captured in $\mathcal{O}(\frac{1}{c})$ or further higher order corrections too. However, this requires the development of new techniques to compute the quantum corrected entanglement entropy. In \cite{Belin:2021htw}, the perturbed area at $O(G_N)$ and the corresponding boundary CFT entanglement entropy of $O(\frac{1}{c})$ have been evaluated. The expression of the bulk entanglement entropy in this order is not available in the literature as per our knowledge. In $h \sim c$ limit, the authors in \cite{Caputa:2022zsr} found the exact expression of the holographic entanglement entropy which is a series in $\frac{h}{c}$. The $O(c^0)$ term in this expression matches with our universal contribution $S_{uni}$ term in CFT or the closed form expression of type-1 minimal area in the bulk. One can check whether the higher order terms in this expansion of $\frac{h}{c}$ shows the signature of dynamical phases too.

 It is important to state here that in the bulk, evolution occurs in time $t$ and not in stroboscopic time $nT$. Thus, as far as the gravity computation is concerned, the role of the drive is to prepare the initial state. i.e: the  entanglement entropy correction is computed after the drive stops. It would be interesting to find the correction to entanglement entropy for the case when the bulk evolution takes place in time $nT$ and not $t$, i.e., during the drive. However, since the entanglement entropy depends only on the state and the choice of subregion, we expect this result to agree with our result for $t=0$. We hope to make progress on this question in the future.

\acknowledgments{We thank  Suchetan Das, Justin David, Somnath Porey, Baishali Roy and Krishnendu Sengupta for useful discussions. SD thanks Pekhom, Ashis, Anirban, Arnab and Amit for their cheerful company from time to time. PD is supported by ANRF Early Career Research Grant ANRF/ECRG/2024/000247/PMS. 
}

\appendix
\section{A few useful results involving the Baker-Campbell-Hausdorff formula}\label{bch}
In this appendix, for completeness, we review some results involving the Baker-Campbell-Hausdorff formula following \cite{Matone:2015wxa}, which we make use in this paper. Given an algebra with two generators $X,Y$, which has the commutation relation:
\begin{align}
    [X,Y]=u X+v Y+c I\,,
\end{align}
where $I$ is the identity operator, and $u,v,c \in \mathbb{C}$,
then one can write:
\begin{align}
    \exp(X) \exp(Y)= \exp\left(X+Y+f(u,v)[X,Y]\right)\,.
\end{align}
The specific form of $f(u,v)$ can be found in \cite{Matone:2015wxa}. Based on this, one can further extend the algorithm, and add another generator $Z$. 
\begin{align}
    [Y,Z]= \omega Y+ z Z+ d I,\, \quad \omega,z, d\in \mathbb{C}\,.
\end{align}
From the Jacobi identity one can write:
\begin{align}
    [Y,Z]=m X+n Y+ p Z+ e I.
\end{align}
Jacobi identity constraints the constants $e,m,n$ and $p$ by the linear equations. One can use the following identity  
\begin{align}
    \exp(X) \exp(Y) \exp(Z)= \exp(X) \exp(\alpha Y) \exp(\beta Y) \exp(Z)\,,\qquad \alpha+\beta=1\,.
\end{align}
Using the above identity, and associativity of the BCH formula one can find the closed form for $W$ as follows.
\begin{align}
    \exp(W)= \exp(X) \exp(Y) \exp(Z),
\end{align}
where $X:= \lambda_{-1} L_{-1}$,~~~$Y:= \lambda_{0} L_{0}$,~~~$Z:= \lambda_{1} L_{1}$, and
\begin{align}
    [X,Y]= \lambda_0 X,~~~[Y,Z]=\lambda_0 Z,~~~[X,Z]=\lambda_{-1}\lambda_{1} \frac{2}{\lambda_0} Y \,.
\end{align}
The closed form is given by
\begin{align}\label{eq: BCH closed}
    &\exp(\lambda_{-1} L_{-1})\exp(\lambda_0 L_0) \exp(\lambda_1 L_1)\nn &= \exp \left \lbrace \frac{\lambda_+ - \lambda_-}{e^{-\lambda_-}-e^{-\lambda_+}} \left[ \lambda_{-1} L_{-1} + (2-e^{-\lambda_+}-e^{-\lambda_-})L_0+ \lambda_1 L_1 \right]\right \rbrace,
\end{align}
where the value of $\lambda_{\pm}$ is given by
\begin{align}\label{eq: BCH seed eq}
    e^{- \lambda_{\pm}}= \frac{1+ e^{-\lambda_0}- \lambda_{-1}\lambda_1 \pm \sqrt{(1+ e^{- \lambda_0}-\lambda_{-1}\lambda_1)^2-4 e^{-\lambda_0}}}{2}\,.
\end{align}
\eqref{eq: BCH closed} can be used to derive the normal ordered form of the prepared state in \eqref{eqn:normalorder}.

\section{$sl(2,\mathbb{R})$ driven CFTs}\label{appendix:drive}
In this section, we will briefly review $sl(2,\mathbb{R})$ driven CFTs \cite{Wen:2018agb, Wen:2020wee, Wen:2018vux, Das:2021gts}. We consider a 2D CFT living on a circle of length $L$ with central charge $c$, evolving under a Floquet Hamiltonian, described by the following two-step discrete drive protocol:
\begin{align}\label{eq: driven Hamiltonian}
H(t)=\begin{cases}
		H_1, & \text{for\,  $0<t<T_1$}\\
            H_0, & \text{for\, $T_1<t<T_1 +T_0$}
		 \end{cases}\,,
         \end{align}
where
\begin{align}\label{eq: driven Hamiltonian2}
& H_0= \frac{2 \pi}{L}\left[ L_0 +\bar{L}_0\right]-\frac{\pi c}{12 L} \,, \\
& H_1= \frac{2 \pi}{L}\left[ L_0 + \bar{L}_0 + \tanh 2\theta \Big(\frac{L_1+L_{-1} +\bar{L}_{1} +\bar{L}_{-1}}{2}\Big) \right]-\frac{\pi c}{12 L}\,.
\end{align}
Here $L_{\pm 1}$ and $L_0$ are the global conformal group generators. $\theta \geq 0$ is a real continuous parameter and $T=T_0+T_1$ is the period of the drive. 
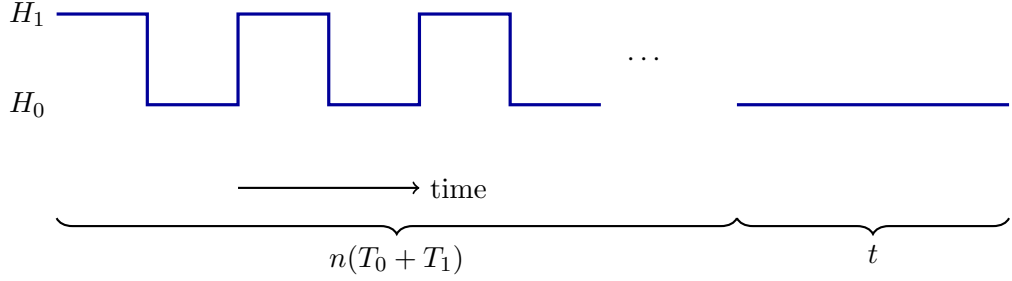
\begin{figure}
\begin{center}
\begin{tikzpicture}[thick]

\def\h{1.5}   
\def\l{0.3}   
\def\w{1.2}   

\draw[blue!60!black, very thick]
(0,\h)
-- (\w,\h)
-- (\w,\l)
-- (2*\w,\l)
-- (2*\w,\h)
-- (3*\w,\h)
-- (3*\w,\l)
-- (4*\w,\l)
-- (4*\w,\h)
-- (5*\w,\h)
-- (5*\w,\l)
-- (6*\w,\l);

\node at (6.5*\w,0.9) {$\cdots$};

\node[left] at (0,\h) {$H_1$};
\node[left] at (0,\l) {$H_0$};

\draw[->, thick] (2*\w,-0.8) -- (4*\w,-0.8) node[right] {time};

\draw[decorate,decoration={brace,mirror,amplitude=6pt}]
(0,-1.2) -- (7.5*\w,-1.2)
node[midway,below=6pt] {$n (T_0+T_1)$};

\draw[decorate,decoration={brace,mirror,amplitude=6pt}]
(7.5*\w,-1.2) -- (10.5*\w,-1.2)
node[midway,below=6pt] {$t$};

\draw[blue!60!black, very thick] (7.5*\w,\l) -- (10.5*\w,\l);

\end{tikzpicture}
\caption{Evolution of the state under the 2-step periodic drive with time period $T_0+ T_1$ for $n$ cycles followed by $H_0$ for time $t > n (T_0+T_1)$. }
\end{center}
\end{figure}
Let us choose the initial state to be the primary state $|h,h \rangle$. We drive the state with $H_1$ for time $T_1$ and then with $H_0$ for time $T_0$, and repeat this for $n$-cycles. The stroboscopic dynamics of the system after $n$-cycle is governed by an effective time-independent Hamiltonian $H_{eff}$, which is given by the equation
\begin{align}
   e^{-i H_{eff} n T}|h,h \rangle \equiv e^{-i H_0 T_0}e^{-i H_1 T_1} \cdots e^{-i H_0 T_0}e^{-i H_1 T_1}|h,h \rangle\,,
\end{align}
and
\begin{equation}
 H_{eff} =\frac{i}{ T}\log\Big(e^{-iH_1T_1}e^{-iH_0T_0}\Big)\,.
\end{equation}
Thus, the Heisenberg evolution of operators in the stroboscopic time $t=nT$, is given by $\mathcal{O}_H(t) =e^{iH_{eff}t}\mathcal{O}(0)e^{-iH_{eff}t}$. Since $H_{eff}$ is the generator of $sl(2,\mathbb{R})$ transformation, it can be expressed as a linear combination of global conformal generators:
\begin{equation}\label{eq: H eff CFT}
H_{eff}=\alpha(L_0+\bar{L}_0)+\beta(L_1+\bar{L}_1)+ \gamma(L_{-1}+\bar{L}_{-1}) . 
\end{equation}
$\alpha$, $\beta$ and $\gamma$ depends on the external parameters $T_0,T_1$ and $\theta$, tuning which, we can induce dynamical phase transitions.

If we concentrate on the holomorphic sector of the Hamiltonian, then we can see the effective Hamiltonian above mixes the primary with all global descendants of the form $|h+m_1,h+m_2 \rangle$, having angular momentum $m=m_1-m_2$. These states have the norm $\mathcal{N}_{m,0}$.
\begin{align}\label{eq: CFT norm}
    \mathcal{N}_{m,0}^2= \langle h,h|L_{1}^m L_{-1}^m|h,h\rangle=\frac{\Gamma(2h+m) m!}{\Gamma(2h)}
\end{align}    
As a further simplification, we can express the evolution operator in a more useful form \cite{Matone:2015wxa, liska:2023scar}, as follows \footnote{See Appendix \ref{bch} for the details.}:
\begin{align}\label{eqn:normalorder}
e^{i H_{eff} nT}  &= e^{A_+^L L_{-1}} e^{\log[A_0^L]L_0}e^{A_-^L L_1},\nn
e^{-i H_{eff} nT} & = e^{A_+^R L_{-1}} e^{\log[A_0^R]L_0}e^{A_-^R L_1}\,,
\end{align}
where
\begin{align}
    & A_{+}^L= \frac{i nT\gamma \frac{\sinh \sig}{\sig}}{\cosh \sig- inT\alpha \frac{\sinh \sig}{2\sig}},\ A_{-}^L= \frac{i T \beta \frac{\sinh \sig}{\sig}}{\cosh \sig- iT\alpha \frac{\sinh \sig}{2\sig}}\,\\
    & \notag A_{+}^R= \frac{-i nT\gamma \frac{\sinh \sig}{\sig}}{\cosh \sig+i nT\alpha \frac{\sinh \sig}{2\sig}},\ A_{-}^R= \frac{-i nT \beta \frac{\sinh \sig}{\sig}}{\cosh \sig+i nT\alpha \frac{\sinh \sig}{2\sig}}\,\\
    & \notag A_0^L=(\cosh \sig-\frac{i nT \alpha}{2\sig} \sinh \sig)^{-2},A_0^R=(\cosh \sig+\frac{i nT \alpha}{2\sig} \sinh \sig)^{-2}\, ,
    \end{align}
with 
\begin{align}\label{eq: def sigma}
\sig=\frac{1}{2} n T (4 \beta \gamma -\alpha^2)^{1/2}\,.
\end{align}

Hence, the normal ordered form of our prepared state becomes
\begin{align}\label{eq: prepared state cft normal order}
    |\hat{\psi}\rangle & = e^{A_+^R L_{-1}} e^{L_0\log[A_0^L]} e^{A_-^R L_1}|h,h\rangle\\
    & \notag = e^{h\log[A_0^R]} \sum_{q=0}^\infty \frac{(A_+^R)^q}{q!} L_{-1}^q |h,h \rangle. 
\end{align}
The expectation value of any observable $\mathcal{O}$ in this state can be simplified as follows,
\begin{align}\label{eq: observable cft}
   & \nonumber \langle \hat{\psi} | \mathcal{O}(0)|\hat{\psi}\rangle\\
   & \nonumber =\langle h,h | e^{A_+^L L_{-1}} e^{\log[A_0^L]L_0}e^{A_-^L L_1} \mathcal{O}(0)
e^{A_+^R L_{-1}} e^{\log[A_0^R]L_0}e^{A_-^RL_1}|h,h\rangle\\
&= e^{h\log A_0^L }e^{h\log A_0^R } \langle h,h | e^{A_-^L L_1} \mathcal{O}(0)e^{A_+^R L_{-1}}|h,h\rangle.
\end{align}

It should be noted that the hermiticity of $H_{eff}$ in \eqref{eq: H eff CFT} demands $\beta=\gamma$. For convenience of the computation, we choose $\alpha=0$, and $\beta,\gamma \in \mathbb{R}$, which yields
\begin{equation}\label{eq: non-heating parameter CFT}
    A_{+}^R=-i \tanh \sigma,\ A_{-}^L=(A_{+}^R)^*,\ A_0^L=A_0^R= (\cosh \sigma)^{-2}\,.
\end{equation}
In section \ref{sec: bulk computation}, we will compute the holographic entanglement entropy of the dual state of \eqref{eq: prepared state cft normal order}, with the above choices of parameters \eqref{eq: non-heating parameter CFT}. The generalization to any arbitrary values of the parameters is straightforward but more involved as we will comment at the end of section \ref{sec: bulk computation}.


\subsection{Details of time evolution under $sl(2,\mathbb{R})$ drive}\label{appendix:mobius}
We now derive the expression for the operator positions as given in equation (\ref{operatorpositions}). 
On the plane, the Euclidean time evolved state ($|\psi(\tau)\rangle$) is given by:
\begin{equation}
|\psi (\tau)\rangle = e^{-\tau H_0}e^{-nTH_{eff}}|h,h\rangle = e^{-\tau H_0}e^{-nTH_{eff}}\mathcal{O}_{h,h}(0,0)e^{nTH_{eff}}e^{\tau H_0}|0\rangle,
\end{equation}
where $H_{eff} = \alpha(L_0 + \bar{L}_0) + \beta(L_{-1} + L_{1} + \bar{L}_{-1} +\bar{L}_{1})$,  and $H_0 = L_0 +\bar{L}_0.$ 
Since $H_{eff}$ generates a conformal transformation, we have :
\begin{equation}
    e^{-nTH_{eff}}\mathcal{O}_{h,h}(0,0)e^{nTH_{eff}} = \Big(\frac{\partial z_n}{\partial z}\frac{\partial \bar{z}_n}{\partial \bar{z}}\Big)^h\Big|_{z=\bar{z}=0} \mathcal{O}(z_n,\bar{z_n}).
\end{equation}
The expression for $z_n$ can be found from the following equation:
\begin{equation}\label{ct}
    e^{n T h_{eff}} z = z_n, 
\end{equation}
where $h_{eff}$ is the space-time representation of the $SL(2, \mathbb{R})$ generators:
\begin{equation}
h_{eff} = -\alpha z\partial z - \beta (z^2 +1)\partial_z +c.c
 \end{equation}
We solve \eqref{ct} separately for the three cases in \eqref{eqn:3phases}. We then map the operator back to the cylinder, and finally analytically continue back to the real time  i.e. $T\rightarrow iT $, and $\tau \rightarrow it$, to obtain the expression for $|\psi(t)\rangle $.

\subsection*{Heating phase}
We first look at the special case where $\alpha =0$ and order parameter $\delta =2\beta$. One way to solve equation (\ref{ct}) is to first make a change of coordinates from $z$ to $\omega$, such that in the new coordinates, $h_{eff}$ generates a translation i.e. $n T h_{eff} \equiv \partial_{\omega}$. Solving for $\omega$, we get
\begin{equation}
    z = -\tan{(nT\beta\omega)}\,.
\end{equation}
Thus, from \eqref{ct}, we get
\begin{equation}
z_n = \tan(-nT\beta + z) = \frac{-\tan(nT\beta) +\tan(z)}{1 +\tan(nT\beta)\tan(z)}\,.
\end{equation}
For $z=0$, one gets $z_n = -\tan(nT\beta)$ and for $z =\infty$, one gets $z_n = \cot{(nT\beta)}$. Further evolving by $e^{\tau H_0}$ gives 
\begin{align}\label{eqn:mob_alpha0}
  z_n(z=0) = -e^{-\tau} \tan(nT\beta)\,,\quad   z_n(z=\infty) = e^{-\tau} \cot(n\beta T)\,.
\end{align}

For the more general case, when $\alpha \neq 0$, the solution of  \eqref{ct} is given by: 
\begin{equation}
    z_n = \frac{|\delta|}{2\beta}\Bigg(\frac{\frac{2\beta}{|\delta|}z + \frac{\alpha}{|\delta|}-\tan\left(\frac{nT|\delta|}{2}\right)}{1+\tan\left(\frac{nT|\delta|}{2}\right)\cdot\left(\frac{2\beta}{|\delta|}z+\frac{\alpha}{|\delta|}\right)} -\frac{\alpha}{|\delta|}\Bigg)\,.
\end{equation}
For the case of $z=0$, we get
\begin{equation}
   z_n = -\frac{2\beta}{|\delta|} \cdot \frac{1}{\Bigg(cot\left(\frac{nT|\delta|}{2}\right) +\frac{\alpha}{|\delta|}\Bigg)} \,,
\end{equation}
while for $z=\infty$, we get 
\begin{equation}
    z_n = -\frac{|\delta|}{2\beta}\Bigg(-\cot\left(\frac{nT|\delta|}{2}\right) + \frac{\alpha}{|\delta|}\Bigg)\,.
\end{equation}
Further evolution by $e^{\tau H_0}$ gives us
\begin{eqnarray}\label{eqn:mob_hp}
    z_n (z=0) &=& \frac{2\beta e^{-\tau}}{|\delta|} \cdot \frac{1}{\Bigg(-\cot(\frac{nT|\delta|}{2}) -\frac{\alpha}{|\delta|}\Bigg)},  \nonumber \\
    z_n (z=\infty) &=& -\frac{|\delta|e^{-\tau}}{2\beta}\Bigg(-\cot(\frac{nT|\delta|}{2}) + \frac{\alpha}{|\delta|}\Bigg).
\end{eqnarray}

\subsection*{Non Heating Phase}
A similar calculation in this case shows that
\begin{align}
z_n &= e^{-2\tau}\Bigg(\frac{|\delta|}{2\beta}\cdot\frac{2z\beta +\alpha +|\delta|\tanh\left(\frac{nT|\delta|}{2}\right)}{|\delta| +(2z\beta +\alpha)\tanh\left(\frac{nT|\delta|}{2}\right)} -\frac{\alpha}{2\beta}\Bigg) \,, 
\end{align}
which gives
\begin{align}\label{eqn:mob_nhp}
z_n (z=0) &= -\frac{2\beta e^{-\tau}}{|\delta|}\cdot\frac{1}{\coth(\frac{nT|\delta|}{2}) +\frac{\alpha}{|\delta|}}\, , \nn
z_{n}(z=\infty) &= -\frac{|\delta|e^{-\tau}}{2\beta}\cdot \Bigg( -\coth(\frac{nT|\delta|}{2}) +\frac{\alpha}{|\delta|}\Bigg) \,.
\end{align}

\subsection*{Phase Boundary} 
In this case, we get
\begin{align}
    z_n &= e^{-\tau}\cdot\Bigg(\frac{2(z+1)}{2 +nT \alpha(z+1)} -1\Bigg)\,,
    \end{align}
which gives
    \begin{align}\label{eqn:mob_pb}
     z_{n} (z=0) &= -e^{-\tau}\cdot\frac{nT\alpha}{2+nT\alpha} \,, \quad
    z_{n}(z=\infty) =e^{-\tau}\cdot\frac{2 - nT\alpha}{nT\alpha}\,.
\end{align}

\section{Details: perturbed minimal Area} \label{appendix: D}

In this section, we provide the key steps of the computation of minimal area. We describe the solution of Einstein's equation, the Fefferman-Graham expansion of the perturbed metric to obtain the integration constants, and finally the computation of the perturbed area due to $r \varphi$ component of the metric. All details of the computation are provided in the accompanying Mathematica notebook.
\subsection{Solution of Einstein's equation}
\label{app: backreaction}
In this section, we state the key steps to solve Einstein's equation in the bulk geometry. The $tt,rr$ and $t\varphi$ components of Einstein's equation are sufficient to determine the perturbed metric. The rest of components serve as a consistency check of the solutions. First, we state the expression of these required stress tensor components for the state $|\hat{\Psi}^{m,n} \rangle $ in \eqref{eq: cft bulk dictionary superposed}.
\begin{align}\label{eq: stress tensor m,n}
    & \langle \hat{\Psi}^{m,n}|T_{{tt}}(t,r,\varphi )|\hat{\Psi}^{m,n}\rangle \\
   \nonumber& = \frac{2}{r^2}\left(r^2+1\right)^{-2 h-m-n+1} \big[ r^{2 m} \left(r^2+1\right)^n \left(2 h (2 h-1) r^2+m^2\right) \left| C_m\right|^2 \mathcal{K}_m^2\\
   & \notag + \left(r^2+1\right)^m r^{2 n}  \left(2 h (2 h-1) r^2+n^2\right) \left|C_n\right|^2 \mathcal{K}_n^2 \big]\\
   & \notag + 2 r^{m+n-2} \left(r^2+1\right)^{-2 h-\frac{m}{2}-\frac{n}{2}+1} \left[2 h (2 h-1) r^2+m n\right] \mathcal{K}_{m}\mathcal{K}_{n} \mathcal{G}_{m,n}(t+\varphi),\\
   & \notag\\
    & \notag \langle \hat{\Psi}^{m,n}|T_{{rr}}(t,r,\varphi )|\hat{\Psi}^{m,n}\rangle\\ 
    \notag & = 4 h (1 + r^2)^{(-1 - 2 h - m - n)} \left[ r^{2 m} \left(r^2+1\right)^n \left| C_m\right|^2 \mathcal{K}_m^2+ \left(r^2+1\right)^m r^{2 n} \left| C_n\right|^2 \mathcal{K}_n^2 \right]\\
    & \notag + 4h r^{m+n} \left(r^2+1\right)^{\frac{1}{2} (-4 h-m-n-2)} \mathcal{K}_{m}\mathcal{K}_{n} \mathcal{G}_{m,n}(t+\varphi), \\
    & \notag \\
    & \notag \langle \hat{\Psi}^{m,n}|T_{t \varphi }(t,r,\varphi )|\hat{\Psi}_{m,n}\rangle\\
    \notag &= (1 + r^2)^{-2 h}\left[2 m (2 h+m)  r^{2 m} \left(r^2+1\right)^{-m} \left| C_m\right|^2 \mathcal{K}_m^2+ 2 n (2 h+n)  r^{2 n} \left(r^2+1\right)^{-n} \left|C_n\right|^2 \mathcal{K}_n^2\right]\\
    \notag & +2 \left[ h (m+n)+m n\right] r^{m+n} \left(r^2+1\right)^{\frac{1}{2} (-4 h-m-n)} \mathcal{K}_{m}\mathcal{K}_{n}\mathcal{G}_{m,n}(t+\varphi).
\end{align}
Note that the expressions are symmetric in $m$ and $n$. We have denoted 
\begin{align}\label{eq: Km def}
\mathcal{K}_{m}= N_{m,0} \mathcal{N}_{m,0}= \frac{1}{\sqrt{2 \pi}} \frac{\Gamma(2h+m)}{\Gamma(2h)}\,,
\end{align}
where  $N_{m,0}$ and $\mathcal{N}_{m,0}$ are, respectively, the norm of the bulk wavefunction $|\Psi_{m,0}\rangle$ and the dual boundary state $|\psi_{m,0}\rangle$ defined in \eqref{eq: normalization const bulk} and \eqref{eq: CFT norm} respectively. The superposition coefficient $C_m$ is defined in  \eqref{eq: Cm general def}. We have also defined a quantity
\begin{align}\label{eq: Gmn def}
    & \mathcal{G}_{m-n}(x) =(C_m^* C_n+ C_m C_n^*)\cos[(m-n)x]+(C_m^* C_n- C_m C_n^*)i\sin[(m-n)x].
\end{align}

Next, we need to find the perturbed metric at  $O(G_N)$ due to the stress tensor components in \eqref{eq: stress tensor m,n} by solving the Einstein's equation \eqref{eq: Einstein}. We assume the following ansatz.
\begin{align}
    ds^2 =&- \left(1+r^2+ G_N  J^{m,n}_{1}(t,r,\varphi)\right) dt^2+ \frac{1}{1+r^2+ G_N J^{m,n}_4(t,r,\varphi)} dr^2+ r^2 d\varphi^2\\
    \notag &+ 2 G_N J^{m,n}_2(t,r,\varphi)\, dt d r + 2 G_N J^{m,n}_3(t,r,\varphi)\,dt d \varphi+ 2 G_N J^{m,n}_5(t,r,\varphi)\,dr d \varphi.
\end{align}
The metric perturbations have the following form.
\begin{align}\label{eq: metric ansatz}
   & J^{m,n}_1(t,r,\varphi)= \mathcal{K}_{m}^2  |C_m|^2 a_m(r)+(m\rightarrow n),\\
   & \notag J^{m,n}_3(t,r,\varphi) =\mathcal{K}_{m}^2 |C_m|^2 b_m(r)+(m\rightarrow n),\\
   & \notag J^{m,n}_4(t,r,\varphi)=\mathcal{K}_{m}^2 |C_m|^2 d_m(r)+(m\rightarrow n),\\
   \notag J^{m,n}_i(t,r,\varphi)= &\mathcal{K}_{m}\mathcal{K}_{n} (m-n)\,\mathcal{G}_{m-n}(t+\varphi+ \frac{3 \pi}{2})\left[ R_{i s}(r)+m n R_{i p}(r)\right],\,\, \text{for}~i=2,5.
\end{align}
$\mathcal{G}_{m,n}(x)$ is defined in \eqref{eq: Gmn def}. The solutions of Einstein's equation are given by

{\small
\begin{align}\label{eq: metric solution}
    & a_m(r)= \tilde{A}_m +\left(r^2+1\right) A_m^\prime+\mathbb{A}_{m}(r),\\
    & \notag b_m(r)= \frac{r^2 B_m^\prime}{2}+\tilde{B}_{m}-8 \pi (2 h+m) r^{2 m+2} \Gamma(m+1) \, \mathcal{F}\left( \left\lbrace m,m,0\right\rbrace,\left\lbrace m,m,2\right\rbrace;\left\lbrace m,m,4\right\rbrace;-r^2 \right),\\
    &  \notag d_m(r) = \tilde{D}_m + 16 \pi r^{2m}\left[\frac{2h}{\left(r^2+1\right)^{2 h+m-1}}-(2 h+m)\, \mathcal{F}\left(\left \lbrace m,m,0 \right\rbrace, \left\lbrace m,m,0\right\rbrace; \left\lbrace m,m,2\right\rbrace;-r^2\right)\right],\\
     & \notag \notag R_{2 p}(r)= R_{5 p}(r)\\
     & \notag = \frac{\tilde{c^p}_{m,n}}{\sqrt{r^2+1}}+\frac{16\pi r^{m+n+1}}{(m-n)^2 (m+n+1)\sqrt{r^2+1}}\mathcal{F}\left(\left\lbrace m,n,1\right\rbrace,\left\lbrace m,n,1\right\rbrace;\left\lbrace m,n,3\right\rbrace ;-r^2\right),\\
    & \notag R_{5 s}(r)= \frac{\tilde{c^s}_{m,n}}{\sqrt{r^2+1}} +\frac{32 \pi h (2 h-1)r^{m+n+3}}{\sqrt{r^2+1} (m-n)^2 (m+n+3)} \, \mathcal{F} \left(\left \lbrace m,n,3\right\rbrace, \left\lbrace m,n,1\right\rbrace;\left\lbrace m,n,5 \right\rbrace;-r^2\right),\\
    & \notag R_{2 s}(r)= \notag R_{5 s}(r)+ \frac{32 \pi \,h\, r^{m+n+1} \left(r^2+1\right)^{\frac{1}{2} (-4 h-m-n)}}{(m-n)^2}.
\end{align}
}
where we have used the following notation
{\small
\begin{align}
_2F_1\left(\ha(p_1+q_1+k_1),\ha(4h+p_2+q_2+k_2);\ha(p_3+q_3+k_3);x\right) \equiv\mathcal{F}\left(\left \lbrace p_1,q_1,k_1  \right \rbrace, \left \lbrace p_2,q_2,k_2\right \rbrace;\left  \lbrace p_3,q_3,k_3\right \rbrace;x\right)
\end{align}
}
The function $\mathbb{A}_{m}(r)$ is the solution of the following equation
\begin{align}\label{eq: Am tt metric}
    \frac{d}{dr} \left[\frac{\mathbb{A}_{m}(r)}{1+r^2}\right]- \frac{2 r}{1+r^2} 16(2h+m) \pi r^{2m} \mathcal{F}\left(\left \lbrace m,m,0\right \rbrace ,\left \lbrace m,m,0\right \rbrace ;\left \lbrace m,m,2\right \rbrace;-r^2 \right)=0.
\end{align}

\subsubsection*{Regularity of metric components at large $r$}
We discuss in some detail the regularity of the metric components \eqref{eq: metric solution}. The second term in \eqref{eq: Am tt metric} vanishes in the leading order in large expansions in $r$ for $h>\ha$. Hence, it is sufficient to choose $A_m^\prime=0$. In $t\varphi$ component, we need to set $ B_{m}^\prime=\frac{16 \pi  \Gamma (2 h+1) \Gamma (m+1)}{\Gamma (2 h+m)}$. With these values the metric components become regular at $r \rightarrow \infty $ for $h>\ha$. The asymptotic values of the metric components are provided below.
{\small
\begin{align}\label{eq: metric large r}
    & a_m(r)|_{r \rightarrow \infty}= d_m(r)|_{r\rightarrow \infty}= \tilde{D}_m-\frac{16 \pi  (2 h+m) \Gamma (2 h) \Gamma (m+1)}{\Gamma(2 h + m)}+O\left( \frac{1}{r}\right) \rightarrow D_m,\\ 
    & \notag b_m(r)|_{r\rightarrow \infty}= \tilde{B}_m+ \frac{8 \pi  m \Gamma (2 h) \Gamma (m+1)}{\Gamma(2h+m)}+O\left( \frac{1}{r}\right) \rightarrow B_m,\\
    & \notag R_{2p}(r)|_{r\rightarrow \infty}=R_{5p}(r)|_{r\rightarrow \infty}= \frac{1}{r} \left[\tilde{c^p}_{m,n}+\frac{8 \pi  \Gamma (2 h) \Gamma \left(\frac{1}{2} (m+n+1)\right)}{(m-n)^2 \Gamma \left(\frac{1}{2} (4 h+m+n+1)\right)}+ O\left( \frac{1}{r}\right) \right] \rightarrow \frac{c^p_{m,n}}{r},\\
    & \notag R_{2s}(r)|_{r \rightarrow \infty}= R_{5s}(r)|_{r\rightarrow \infty}=\frac{1}{r} \left[\tilde{c^s}_{m,n}+ \frac{8 \pi  \Gamma (2 h+1) \Gamma \left(\frac{1}{2} (m+n+3)\right)}{(m-n)^2 \Gamma \left(\frac{1}{2} (4 h+m+n+1)\right)} +O\left( \frac{1}{r}\right) \right] \rightarrow \frac{c^s_{m,n}}{r}.
\end{align}
}
The constants $A_m, D_m, c^s_{m,n}, c^p_{m,n}$ are fixed by comparing the $tt$ component of the holographic stress tensor in \eqref{eq: FG stress tt full}  with the one obtained in the boundary CFT \eqref{eq: boundary stress tt full}. We state their expressions below.
\begin{align}\label{eq: integration const}
   & A_m=D_m=-8(2h+m) \frac{\mathcal{N}_{m,0}^2}{\mathcal{K}_m^2},\\
   & \notag  (m n c^p_{m,n}+ c^s_{m,n} )= 4 (-1)^{m-n}\frac{[h (m-n+1)+n]}{(m-n)^2 \mathcal{K}_m \mathcal{K}_n} \frac{\Gamma (2 h+n)m!}{\Gamma (2 h)}.
\end{align}
The constant $B_m$ can be obtained from the $t \varphi$ component of the stress tensor, but it does not contribute in the area at a constant time slice. Hence, we do not need to evaluate it in this paper.

\subsubsection*{Fefferman-Graham Expansion}
\label{App: FG expansion}
 In an asymptotically $AdS_3$ bulk spacetime $(z,x_i)$, the metric in the Fefferman-Graham form is given by 
\begin{align}\label{FG:coordinate}
ds^2= \frac{1}{z^2}\left( dz^2 + g_{ij}(x,z) dx^i dx^j \right).
\end{align} 
$z=0$ is the boundary. The metric $g(x,z)$ is expanded as
\begin{equation}
g(x,z)= g_{(0)}+ z^2 g_{(2)}+ \cdots.
\end{equation}
Once the metric is in this form, we can read out the expectation value of the stress tensor in the CFT using the expression given below\cite{Balasubramanian:1999re,deHaro:2000vlm}\footnote{There is a difference in a factor of $2\pi$ compared to \cite{deHaro:2000vlm} to take into account the circumference of the cylinder. }.
\begin{align}\label{eq: holographic stress}
\langle T_{ij} \rangle_{ FG} = \frac{1}{4 G_N} g_{(2)ij}.
\end{align}

First, we find the $tt$ component holographic stress tensor for the superposed state $|\hat{\Psi}^{m,n} \rangle= C_m |\Psi_{m,0} \rangle + C_n |\Psi_{n,0} \rangle$. We use the following coordinate transformation in the asymptotic form of the metric perturbation due to the excited state $|\hat{\Psi}^{m,n} \rangle$ (the asymptotic metric can be found by substituting \eqref{eq: metric large r} in \eqref{eq: metric ansatz}).
 \begin{align}
t \rightarrow t + \beta(t,\varphi) z^2,\qquad  r \rightarrow \frac{1}{z+ \alpha (t,\varphi)~z^3}, \qquad \varphi \rightarrow \varphi+\gamma(t,\varphi) z^2 \, ,
\end{align}
where we need to determine $\alpha, \beta, \gamma$. Expanding the metric coefficients as a power series  in $z$ we demand the following conditions so that the metric reduces to the Fefferman-Graham form.
\begin{enumerate}
\item
The $z^{-1}$ term in the $zz$ component of the metric should vanish,
\item
The $zt$ and $z\varphi$ components of the metric vanishes at the leading order. 
\end{enumerate}
These conditions yield the values for $\alpha(t, \varphi), \beta(t, \varphi)$ and $\gamma(t, \varphi)$. From the coefficient of $z^2$ in $tt$ and $\varphi\varphi$ components of the metric, we can compute the respective components of the stress tensor. Further, imposing the condition of tracelessness we obtain
\begin{align}
   A_m= D_m.
\end{align}
Finally, we obtain the following expression of the $tt$ component of the stress tensor 
\begin{align}\label{eq: FG stress tt}
    4 G_N \langle \hat{\Psi}^{m,n} | T_{tt}| \hat{\Psi}^{m,n} \rangle_{FG}&= -\frac{1}{2}-\frac{1}{2}G_N \left( \mathcal{K}_m^2 D_m \left|C_m\right|^2 +D_n \mathcal{K}_n^2 \left|C_n\right|^2\right)\\
    &\notag +G_N (m-n)^2(c^s_{m,n}+mn c^p_{m,n}) \mathcal{K}_{m}\mathcal{K}_{n}\mathcal{G}_{m,n}(t+\varphi).
\end{align}
$\mathcal{G}_{m,n}(x)$ has been defined in \eqref{eq: Gmn def}. The first term $-\frac{1}{2}$ should match with the contribution from the central charge in the boundary stress tensor. We can write down the $tt$ component of the holographic stress energy tensor for the state $|\hat{\Psi}\rangle$
\begin{align}\label{eq: FG stress tt full}
\notag \frac{\langle \hat{\Psi}|T_{tt}|\hat{\Psi}\rangle_{FG}}{\langle \hat{\Psi}|\hat{\Psi}\rangle} & = -\frac{1}{\sum_{p=0} |C_p|^2 \mathcal{N}_{p,0}^2}\bigg[\frac{1}{8 G_N}+\frac{1}{8}\sum_{m,n=0}^{\infty}\left(\mathcal{K}_m^2 D_m \left|C_m\right|^2+D_n \mathcal{K}_n^2 \left|C_n\right|^2\right)\\
& -\frac{1}{4}\sum_{m,n=0,m>n}^{\infty} (c^p_{m,n} m n+c^s_{m,n}) (m-n)^2 \mathcal{K}_{m}\mathcal{K}_{n}\mathcal{G}_{m,n}(t+\varphi)\bigg].
\end{align}

We can also compute the $t \varphi$ component of the stress tensor in a similar way.

\subsubsection*{Boundary stress tensor}
\label{App: boundary stress tensor}
In this subsection, we compute the $tt$ component of the stress tensor of the boundary state $| \hat{\psi}\rangle$ in \eqref{eq: prepared state cft normal order main}. 
First, we find it for the superposed state $|\hat{\psi}_{m,n} \rangle$ in \eqref{eq: cft bulk dictionary superposed}.
\begin{align}\label{eq: boundary stress formula}
   & \langle \hat{\psi}^{m,n}| \mathcal{T}_{tt} | \hat{\psi}^{m,n}\rangle\\
    =& \nonumber  (C_m^*  \langle h+m|+ C_n^* \langle h+n|) \left\lbrace \sum_{k=-\infty}^{\infty} L_k e^{i k (t+\varphi)} +\bar{L}_k e^{i k (t-\varphi)}\right\rbrace(C_m |h+m\rangle +C_n| h+n\rangle),
\end{align}
where we have used the following shorthand notation for states
\[| h+p\rangle \equiv L_{-1}^p| h,h \rangle,\]
whose norm is $\mathcal{N}_{p,0}^2 \equiv \langle h+p| h+p \rangle= \frac{\Gamma(2h+p)p!}{\Gamma(2h)}$. Using \eqref{eq: boundary stress formula} and BCH formula \eqref{eq: observable cft}, we can obtain the stress tensor for $|\hat{\psi}\rangle$ in the holomorphic sector. We provide the final expression below.
\begin{align}\label{eq: boundary stress tt full}
    & \frac{\langle \hat{\psi}| \mathcal{T}_{tt} | \hat{\psi} \rangle}{\langle \hat{\psi}|\hat{\psi} \rangle}\\
    \notag & = \frac{1}{\sum_{p=0}^\infty |C_p |^2 \mathcal{N}_{p,0}^2}\bigg[ \sum_{m=0}^\infty |C_m |^2 (2h+m) \mathcal{N}_{m,0}^2 \\
    & \notag +\sum_{m,n=0, m>n}^\infty  [h (m-n+1)+n] \frac{\Gamma (2 h+n)m!}{\Gamma (2 h)}\,\mathcal{G}_{m,n}(-t-\varphi)\bigg]
\end{align}
Note that the relative sign of $\cos(t+\varphi)$ and $\sin(t+\varphi)$ is different from \eqref{eq: FG stress tt full}. This contributes to the difference of sign in $A_m, D_m$ in \eqref{eq: integration const} for even and odd $m-n$. We can sum over $m,n$ to obtain the total energy
{\small
\begin{align}
    \frac{\langle \hat{\psi}| \mathcal{T}_{tt} | \hat{\psi} \rangle}{\langle \hat{\psi}|\hat{\psi} \rangle} =\frac{2 h \left(\tanh ^4(\sigma )+\tanh ^2 \sigma-\tanh \sigma \left(2 \left(\tanh ^2\sigma +1\right) \sin (t+\varphi )+\tanh \sigma \cos (2 (t+\varphi))\right)+1\right)}{\left(\tanh ^2 \sigma -2 \tanh \sigma \sin (t+\varphi )+1\right)^2}
\end{align}
}
We have plotted $\mathcal{T}_{tt}$ as a function of $\varphi$ at $t=0$ in fig.\ref{fig: bulkplot}.

Next, we describe a bit on how to derive \eqref{eq: boundary stress tt full}. Its first term on the RHS is easy to compute. The $m>n$ terms in the second line can be found as follows. First, we note the following identities.
{\small
\begin{align}\label{eq: boundary stress tensor identities}
   p \geq 1:~~ &\langle h| L_p L_{-1}^p|h \rangle= \frac{(p+1)!}{2} \langle h | L_1 L_{-1} | h \rangle,\\
   \notag p \geq 2:~~ & \langle h | L_{1}L_{p-1}L_{-1}^p|h \rangle=\frac{p!}{2}  \left[ \langle h | L_1^2 L_{-1}^2| h \rangle + 2h(p-2)\langle h | L_1 L_{-1}| h\rangle \right],\\
   \notag p \geq 3:~~ &\langle h | L_1^2 L_{p-2}L_{-1}^p | h \rangle\\
   & \notag = \frac{(p-1)!}{2} \left \lbrace \langle h | L_1^3 L_{-1}^3 | h \rangle + (4h+2)\left[(p-3) \langle h | L_1^2 L_{-1}^2 | h \rangle + \frac{(p-2)(p-3)}{2} 2h \langle h | L_1 L_{-1} | h \rangle \right]\right \rbrace,\\
   \notag p \geq 4:~~ &\langle h| L_1^3 L_{p-3} L_{-1}^p| h \rangle\\
   & \notag  = \frac{(p-2)!}{2}\bigg\lbrace \langle h | L_1^4 L_{-1}^4 | h \rangle + (6h+6)\bigg[(p-4) \langle h |L_1^3 L_{-1}^3 | h \rangle +(4h+2)\frac{(p-4)(p-3)}{2} \langle h | L_1^2 L_{-1}^2 | h \rangle \\
   & \notag  +(4h+2)(2h)\frac{(p-4)(p-3)(p-2)}{6} \langle h | L_1 \Lm | h \rangle \bigg]\bigg \rbrace.
\end{align}
}
Each identity can be found iteratively. We demonstrate the identity for $p \geq 3$ below (here we have suppressed the state $|h\rangle$ for brevity).
{\small
\begin{align} \label{eq: expression L^2 Lp-2 Lm^p}
     L_1^2 L_{p-2}L_{-1}^p & =  (p-1) L_1^2 L_{p-3}L_{-1}^{p-1}+ L_1^2 L_{-1} L_{p-2}L_{-1}^{p-1}\\
    & = \nonumber (p-1) L_1^2 L_{p-3}L_{-1}^{p-1}+ (4h+2) L_1 L_{p-2}L_{-1}^{p-1} \\
    & = \nonumber (p-1)\left \lbrace L_1^2 L_{p-4} L_{-1}^{p-2}+ (4h+2) L_1 L_{p-2} L_{-1}^{p-1}\right \rbrace+ (4h+2)L_1 L_{p-1}L_{-1}^p\\
    & \nonumber \vdots\\
    &= \nonumber \frac{(p-1)!}{2}  \left \lbrace L_1^3 L_{-1}^3+ (4h+2)\left[(p-3) L_1^2 L_{-1}^2 + (1+2+\dots + (p-3))2h L_1 L_{-1}\right]\right \rbrace \\
    & =\nonumber \frac{(p-1)!}{2} \left \lbrace L_1^3 L_{-1}^3+ (4h+2)\left[(p-3) L_1^2 L_{-1}^2 + \frac{(p-2)(p-3)}{2} 2h L_1 L_{-1}\right]\right \rbrace.
\end{align}
}
Using this, we write down the following component of the stress tensor ($p \geq 3$).
\begin{align}\label{eq: boundary stress C2 Cp}
    & C_2^* C_p \langle h+2|\mathcal{T}_{tt}|h+p\rangle + C_p^* C_2 \langle h+p| \mathcal{T}_{tt}|h+2\rangle\\
    & \nonumber = C_2^* C_p \langle h| \left(L_1^2 L_{p-2} L_{-1}^p e^{i(p-2) (t+\varphi)}\right)|h\rangle + C_p^* C_2 \langle h| \left( L_{1}^p L_{-p+2} L_{-1}^2e^{-i(p-2) (t+\varphi)}\right)|h\rangle\\
    & \nonumber = \frac{(p-1)!}{2} \left \lbrace L_1^3 L_{-1}^3+ (4h+2)(p-3) L_1^2 L_{-1}^2 + 2h(4h+2)\frac{(p-2)(p-3)}{2} L_1 L_{-1}\right \rbrace\\
    & \notag \times \left\lbrace(C_2^* C_p+C_p^* C_2)\cos[(p-2)(t+\varphi)]-i(C_p^* C_2-C_2^* C_p)\sin [(p-2)(t+\varphi)]\right\rbrace
\end{align}
Finally, the expectation value of the boundary stress tensor for arbitrary $m$ and $n$ ($m>n$) can be evaluated as
\begin{align}\label{eq: stress tensor comp general m,n}
& C_m^* C_n \langle h+m|\mathcal{T}_{tt}|h+n\rangle + C_n^* C_m \langle h+n|\mathcal{T}_{tt}|h+m\rangle\\
\notag &= [h (m-n+1)+n] \frac{\Gamma (2h+n)m!}{\Gamma (2 h)}\\
\notag &\times \left[(C_m^* C_n+C_m^* C_n)\cos[(m-n)(t+\varphi)]-i(C_m^* C_n-C_n^* C_m)\sin [(m-n)(t+\varphi)]\right].
\end{align}

\subsection{Perturbed area: $r \varphi$ component}
\label{App: details area}
In this section, we give details of the computation of the minimal area corresponding to the $r\varphi$ component of the perturbed metric.

For a specific $m$ and $n$, this area can be obtained by substituting $\delta g_{r \varphi}$, \eqref{eq: metric non-heating} in the first and second term  of the expression of the area in \eqref{eq: area formula}.
\begin{align}\label{eq: area r varphi NH}
    & \delta A_{r\varphi}(t)=\frac{1}{\hat{N}}\sum_{m,n=0}^\infty(m-n)\frac{(-i n T \gamma)^{m+n}}{m! n!} \mathcal{K}_m \mathcal{K}_n\\
    & \times \nonumber \int^{r_{min}}_\infty dr~(R_{5s}(r)+ mn R_{5p}(r)) \frac{r_{min}}{r^2} (\mathcal{G}_{m,n}^{0}(t+\varphi_I)+\mathcal{G}^0_{m,n}(t+\varphi_{II}))\\
    & \notag \equiv \frac{1}{\hat{N}}G_N(m-n)\frac{(-i n T \gamma)^{m+n}}{m! n!} \mathcal{K}_m \mathcal{K}_n \sum_{i=1}^3 \mathcal{I}_{m,n;i}(r_{min},t).
\end{align}
The function $\mathcal{G}_{m,n}^{0}(x)$ is defined in \eqref{eq: def G0}. The index $i$ in $\mathcal{I}$ corresponds to the three types of the metric. We first concentrate on the type 1 term which is proportional to $ c_s^{m,n}+m n c_p^{m,n}$. The type-1 area due to the states with $m-n=1$ does not follow any pattern, hence we evaluate this first. 
{\small
\begin{align}\label{eq: area r varphi m-n=1 type 1}
    \notag \delta_1 A_{r\varphi}(t) &=- \frac{1}{\hat{N}} (m-n)\frac{(-i nT \gamma)^{m+n}}{m! n!} \mathcal{K}_m \mathcal{K}_n (c_p^{m,n} m n+c_s^{m,n})\left[(-1)^{m}-(-1)^n\right]\\
    & \nonumber \times \bigg\lbrace\int^{r_{min}}_\infty d r ~\frac{r_{min}}{r^2}\frac{i\cos [(t+\varphi_I)]}{\sqrt{r^2+1}}+\int^{r_{min}}_{\infty} d r ~\frac{r_{min}}{r^2} \frac{i\cos [(t+\varphi_{II})]}{\sqrt{r^2+1}}\bigg \rbrace\\
    & \notag =- \frac{1}{\hat{N}}(m-n)\frac{(-i n T \gamma)^{m+n}}{m! n!} \mathcal{K}_m \mathcal{K}_n (mn ~c_p^{m,n}+c_s^{m,n})\left[(-1)^{m}-(-1)^n\right]\\ 
    & \times \left(\cos \pi x -\frac{\pi x}{\sin \pi x}\right)\left\lbrace i \sin(t+\pi x)\right \rbrace, 
\end{align}
}

where we have used the following identities:
\begin{align*}
    & \rm=\frac{1}{2 i}\left[(-1+ i \rm)+(1+ i \rm)\right],\\
    & 1= \frac{1}{2}\left[(1+ i \rm)-(-1+ i \rm)\right],
\end{align*}
and the relation between the half interval $\pi x$ and $r_{min}$ \eqref{eq: pix and rm relation}. 
For states with $m-n>1$ we find the following.
\begin{align}
    \mathcal{I}_{m,n;1}(r_{min},t)= & (c_p^{m,n} m n+c_s^{m,n})\left[(-1)^{m}-(-1)^n\right] \times I_{m-n}(r_{min},t), \text{for}~(m-n)=\text{odd},\\ 
    = \notag  & G_N (c_p^{m,n} m n+c_s^{m,n})\left[(-1)^{m}+(-1)^n\right] \times I_{m-n}(r_{min},t), \text{for}~(m-n)=\text{even},
\end{align}

where we have separated out the $r_{min}$ dependent part $I_{m-n}(r_{min},t)$. $I_{m-n}(r_{min},t)$ can be expressed in a convenient form using  the quantities $(1+i r_{min})^{m-n}$ and $(-1+i r_{min})^{m-n}$ and the relation between $\pi x$ and $r_m$ \eqref{eq: pix and rm relation}.

The final expression  of $I_{m-n}(r_{min})$ is given by
\begin{align}\label{eq: I(rm) expression}
    I_{m-n}(r_{min},t) \notag & = \frac{(m-n)}{(m-n-1)(m-n+1)} \mathcal{E}_{m-n}(x) \left \lbrace 2 i \sin((m-n)(t+\pi x))\right \rbrace,~~ \text{for}~m-n=\text{odd},\\
    & =-\frac{(m-n)}{(m-n-1)(m-n+1)} \mathcal{E}_{m-n}(x) \left \lbrace 2 \cos((m-n)(t+\pi x))\right \rbrace,~~ \text{for}~m-n=\text{even},
\end{align}
where we have defined the following function 
\begin{align}\label{eq: E(m-n)(x)}
    \mathcal{E}_{m-n}(x)= \left( \cos ((m-n)\pi x)-\frac{\cot \pi x}{m-n} \sin ((m-n)\pi x) \right). 
\end{align}
We show the example for $m-n=4$ below.

\paragraph{m-n=4:}~\\
$I_{4}(r_{min},t)$ can be computed as
\begin{align}
    I_4(r_{min},t)=- \left(5 r_{min}^2-1\right) \left \lbrace \sin 4t \frac{32 r_{min} \left(r_{min}^2-1\right) }{15 \left(r_{min}^2+1\right)^4} -\cos 4t \frac{8 \left(r_{min}^4-6 r_{min}^2+1\right)}{15 \left(r_{min}^2+1\right)^4}\right \rbrace.
\end{align}
We use the following identities to simplify the above expression.
\begin{align*}
    &  (5 r_{min}^2-1)=-\left(\frac{i}{\sin\pi x}\right)^{4}\mathcal{E}_{4}(x),\\
     & (-1+i r_{min})^4+(1+i r_{min})^4=2 r_{min}^4-12 r_{min}^2+2,\\
    & (1+i r_{min})^4-(-1+i r_{min})^4= 8 i r_{min}-8 i r_{min}^3.\\
    \end{align*}
Finally, we obtain
{\small
\begin{align}
  \notag I_4(r_{min},t) & =-\frac{4}{15}(5r_{min}^2-1)\\
  \notag & \times\left\lbrace i \sin 4t \frac{\left[(1+i r_{min})^4-(-1+i r_{min})^4\right]}{\left(i r_{min}+1\right)^4\left(1-i r_{min}\right)^4}-\cos 4t \frac{\left[(1+i r_{min})^4+(-1+i r_{min})^4\right]}{\left(i r_{min}+1\right)^4\left(1-i r_{min}\right)^4}\right\rbrace\\
  &\nonumber =-\frac{4}{3 \times 5} \mathcal{E}_4(x)\left \lbrace 2 \cos 4(t+\pi x) \right \rbrace.
\end{align}
}
For convenience of readers, we state the following useful identities which we have used in our calculation ($\tilde{\mathcal{E}}_{m-n}(x)= \left(\frac{i}{\sin \pi x}\right)^{m-n}\mathcal{E}_{m-n}(x)$).
{\small
\begin{align}
    & \notag m-n=2: 1= \tilde{\mathcal{E}}_{2}(x),~m-n=6: 35 r_{min}^4-42 r_{min}^2+3=3 \tilde{\mathcal{E}}_3(x),\\
    & \notag m-n=8: 21 r_{min}^6-63 r_{min}^4+ 27 r_{min}^2-1=- \tilde{\mathcal{E}}_8(x),\\
    & \notag m-n=3: i r_{min} =-\frac{3}{8} \left(\frac{i}{\sin\pi x}\right)^{3}\mathcal{E}_{3}(x),~m-n=5: \frac{24}{5}\left(\frac{5 r_{min}^2}{3}-1 \right) i r_{min}= - \tilde{\mathcal{E}}_5(\pi x),\\
    & \notag m-n=7 : 48 \left[ \frac{7}{3} (r_{min}^2-2)r_{min}^2+1 \right]= \frac{7}{i r_{min}} \tilde{\mathcal{E}}_7(x),\\
    & m-n=9: 80(-1+ 9 r_{min}^2-\frac{63}{5} r_{min}^2+ 3 r_{min}^6)=-\frac{9}{i r_{min}} \tilde{\mathcal{E}}_9(x).
\end{align}
}

The final expression of the area due to the type 1 term of metric $\delta g_{r \varphi}$ can be written down from \eqref{eq: area r varphi NH}, \eqref{eq: area r varphi m-n=1 type 1}, and \eqref{eq: I(rm) expression}.
{\small
\begin{align}\label{eq: area r varphi NH type1}
   \delta_1 A_{r\varphi} = \notag & \frac{1}{\hat{N}} \bigg \lbrace -\sum_{m,n=0,(m-n)=1}^\infty \mathcal{K}_m \mathcal{K}_n(m-n)^2(mn c_p^{m,n}+c_s^{m,n})\frac{(-inT \gamma)^{m+n}}{m! n!}\left[(-1)^{m}-(-1)^n\right] \\
    & \notag \times i\sin [(t+\pi x)] \left[\frac{\pi x}{\sin\pi x}-\cos\pi x\right]\\
   & \notag- \sum_{m,n=0, \text{even}~(m-n)}^\infty \mathcal{K}_m \mathcal{K}_n(m-n)^2(mn c_p^{m,n}+c_s^{m,n})\frac{(-in T \gamma)^{m+n}}{m! n!}\frac{ \left[(-1)^{m}+(-1)^n\right]}{(m-n-1)(m-n+1)}\\
    \nonumber & \times 2\cos [(m-n)(t+\pi x)]\left[\cos (m-n)\pi x-\cot \pi x \frac{\sin (m-n)\pi x}{m-n}\right] \\
    & \notag +\sum_{m,n=0, \text{odd}~(m-n)}^\infty \mathcal{K}_m \mathcal{K}_n(m-n)^2(mn c_p^{m,n}+c_s^{m,n})\frac{(-inT \gamma)^{m+n}}{m! n!}\frac{ \left[(-1)^{m}-(-1)^n\right]}{(m-n-1)(m-n+1)}\\
    & \notag \times 2i \sin[(m-n)(t+\pi x)]\left[\cos (m-n)\pi x-\cot \pi x \frac{\sin (m-n)\pi x}{m-n}\right] \bigg \rbrace\\
    & \equiv  A_{r\varphi}|_1 + A_{r\varphi}|_{>1},
\end{align}
}
where we denote the sum with constraint $m-n=1$ as $A_{r\varphi}|_1$, and $m-n>1$ as $A_{r\varphi}|_{>1}$.  The value of $(mn c_p^{m,n}+c_s^{m,n})$ is given in \eqref{eq: integration const}. We have performed the sum in the Mathematica. The details are provided in the accompanying notebook. We state the final expression here.
\begin{align}
    & A_{r\varphi}|_{1}= -\frac{8 G_N h \gamma n T (\sin 2 \pi  x-2 \pi  x) \csc \pi  x \sin (t+\pi  x)}{\gamma ^2 n^2 T^2-1}\\
    & \notag A_{r\varphi}|_{>1}= \frac{4 G_N h}{\gamma ^2 n^2 T^2-1} \bigg[ 4 \gamma ^2 n^2 T^2 +4 \gamma  n T \cos \pi  x \sin (t+\pi  x) \\
    & \notag +i \cot \pi  x \left(2 \gamma n T \sin t+\gamma ^2 n^2 T^2+1 +2 \gamma n T \cos t\right)\left( L(t+2 \pi x)-L(t) \right)\bigg],
    \end{align}
where we have defined $L(t)=\log \frac{\left(1-i \gamma nT e^{i t} \right)}{\left(1+i \gamma nT  e^{-i t} \right)}$.

\par\noindent\rule{0.3 \textwidth}{0.5 pt}

Next, we consider the type 2 term. We note the corresponding metric component from \eqref{eq: metric non-heating},
\begin{align}
 \delta_2 g_{r\varphi}(r)= -\frac{16 \pi h r^{m+n+1}(1+r^2)^{\frac{1}{2} (-4 h-m-n)}}{(m-n)^2},
\end{align}
and substitute in the expression of area \eqref{eq: area r varphi NH} with $i=2$. We provide the expression of the leading order term of $\mathcal{I}_{m,n;2}(r_{min},t)$ in large $r_{min}$ expansion.
\begin{align}\label{eq: Imn general}
    \mathcal{I}_{m,n;2} = & -\cos[(m-n)t] [(-1)^m+(-1)^n] r_{min}^{-4 h}\frac{8\left(h \pi ^{3/2} \Gamma (2
   h)\right)}{(m-n) \Gamma \left(2 h+\frac{3}{2}\right)}+ O \left( \frac{1}{r^{4h+1}}\right),~ m-n=even,\\
   = \notag & i \sin[(m-n)t] [(-1)^m-(-1)^n] r_{min}^{-4 h}\frac{8\left(h \pi ^{3/2} \Gamma (2
   h)\right)}{(m-n) \Gamma \left(2 h+\frac{3}{2}\right)}+ O \left( \frac{1}{r^{4h+1}} \right) ,~ m-n=odd.
\end{align}
This can be derived noting the area for a few values of $m-n$. The final expression for the area at the leading order in $O((\pi x)^{4h})$ becomes
\begin{align}\label{eq: area r varphi type 2}
   \delta_2 A_{r\varphi} & = - \frac{1}{\hat{N}}  G_N (\pi x)^{4 h}\frac{8 h \pi ^{3/2} \Gamma (2
   h)}{\Gamma \left(2 h+\frac{3}{2}\right)}  \\
   & \notag \bigg\lbrace \sum_{m,n=0,m-n=odd}^\infty  -i \frac{(-i n T\gamma)^{m+n}}{m! n!} [(-1)^m-(-1)^n] \sin[(m-n)t] \mathcal{K}_{m}\mathcal{K}_{n}\\
   & \notag + \sum_{m,n=0,m-n=even}^\infty\frac{(-i n T\gamma)^{m+n}}{m! n!} [(-1)^m+(-1)^n] \cos[(m-n)t]\mathcal{K}_{m}\mathcal{K}_{n}\bigg \rbrace.
\end{align}

\section{\bbgv coefficients}
\label{app: bbgv coefficients}

We would like to find the \bbgv coefficients for the bulk state $|\Psi_{m,0} \rangle$ in the short distance approximation following \cite{Belin:2018juv,Chowdhury:2024fpd}. First, we find the near boundary behavior of the following two point function.
\begin{equation}\label{eq:bbgv-2-pt-func m0}
\lim_{r \to \infty} r^{2 h} \, \langle 0| \phi(r,t,\varphi) \, a^\dagger_{m,0} \, |0\rangle = N_{m,0}e^{-i(2h+|m|)t}e^{-i m \varphi}  \, .
\end{equation}
$N_{m,0}$ is defined in \eqref{eq: normalization const bulk}. The strategy is to write the field $\phi$ in terms of the Rindler coordinates and extract the expression of the Bogoliubov coefficients from the above two point function. The \bbgv coefficients of the descendants scales w.r.t that of the primary excitation \cite{Chowdhury:2024fpd}. In the short distance approximation, using the AdS-Rindler map in \eqref{coordchange}, it is straightforward to show that the \bbgv coefficients at time $t^\prime$ have the following scaling relation \cite{Chowdhury:2024fpd}.
\begin{align}\label{eq:bbgv scaling}
& \lim_{\eta \rightarrow \infty}\frac{\alpha_{m,0; \om,k}}{\alpha_{0,0; \om,k}}= \frac{N_{m,0}}{N_{0,0}} e^{im (t'+\pi x-2 \pi a)},\\
\notag & \lim_{\eta \rightarrow \infty}\frac{\beta_{m,0; \om,k}}{\beta_{0,0; \om,k}}= \frac{N_{m,0}}{N_{0,0}} e^{-im (t'+\pi x-2 \pi a)}.
\end{align}
We have replaced $|m|$ with $m$ above as we have only used the $m>0$ modes in our computation. In short distance approximation, it is sufficient to know the leading order term of the \bbgv coefficients in the large $\eta$ limit  for the primary excitation. These are given by \cite{Belin:2018juv}
\begin{align}
    \alpha_{0,0;\om,k}=-\beta_{0,0;\om,k} =\frac{2^{2h} (\cosh \eta)^{-2h}}{N_{\om,k}^*\sqrt{8 \pi} \Gamma(2h)^2} \left| \Gamma\left( h+ \frac{i(k-\om)}{2}\right) \Gamma\left( h+ \frac{i(k+\om)}{2}\right)\right|^2,
\end{align}
where $N_{\omega,k}$ is the normalization constant of the Rindler wavefunctions, given in \eqref{eq: Rindler normalization} .

\section{Comparison with previous results}\label{appendix I}
In \cite{Belin:2018juv}, the authors computed the entanglement entropy of the generalised coherent states in 2d CFT using Banados geometry, which is related to the boundary state we considered in \eqref{eq: prepared state cft normal order main} with $\alpha=0$ and holomorphic drive (for whom we have also computed the entanglement entropy in the bulk) . In their notation, we use $k=1$ to obtain our boundary state $|\hat{\psi}\rangle$ \eqref{eq: prepared state cft normal order main}.
\begin{align}
    | \Psi(\xi) \rangle=e^{\xi L_{-1}-\bar{\xi}L_1}|h,h\rangle\,.
\end{align}
The complex coordinate $\xi$ and its complex conjugate $\bar{\xi}$ characterize the coherent states. It is convenient to parametrize the complex coordinate as $\xi=\rho e^{i\theta}$. For an interval $(z_1,\bar{z}_1)$ and $(z_2,\bar{z}_2)$, the regularised or vacuum-subtracted holographic entanglement entropy can be written as
\begin{align}\label{eqn:ee_caputa}
    S=\frac{c}{6}\bigg(\log \bigg|\frac{(f(z_1)-f(z_2))^2}{\epsilon_{UV}^2 f'(z_1)f'(z_2)}\bigg|+\log \bigg|\frac{(\bar{f}(\bar{z}_1)-\bar{f}(\bar{z}_2))^2}{\epsilon_{UV}^2 \bar{f'}(\bar{z}_1)\bar{f'}(\bar{z}_2)}\bigg|\bigg)\,,
\end{align}
$\epsilon_{UV}$ denotes the UV cutoff of the boundary theory. In bulk dual, this corresponds to the holographic cutoff. The functions $f(z)$ and $\bar{f}(\bar{z})$ can be obtained from solving the uniformization equation which results in
\begin{align}
    f(z)&=\bigg(\frac{z-y}{z-\frac{1}{\overline{y}}}\bigg)^{\sigma}\,,\qquad
     \bar{f}(\bar{z})= \bar{z}^{\sigma}\,,
\end{align}
where 
\begin{align}\label{eqn:y}
 y=e^{i \theta} \tanh \rho\,,\quad \sigma=\sqrt{1-\frac{24 h}{c}}\,.  
\end{align}
The endpoints of the interval $(e^{2 \pi a},e^{2 \pi a+ 2\pi x})$ are at $z_1=e^{2 \pi a}$ and $z_2= e^{2 \pi a+ 2\pi i x}$, for $0 \leq a \leq 1$. For simplicity, we assume $a=0$ here. The equation \eqref{eqn:ee_caputa}, being a series in $\frac{h}{c}$ gives entanglement entropy for $h \sim c$. However, the term $O(c^0)$ corresponds to the universal part of entanglement entropy in \eqref{eqn:cftuni}. We show this below.

Let us now compute \eqref{eqn:ee_caputa} for the following class of states 
\begin{align}\label{eqn:coherent_state_caputa}
  e^{-i t L_0}e^{-i n T \beta (L_{-1}+L_1)}|h,h\rangle   \equiv e^{-i t L_0}| \Psi(\xi) \rangle\,.
\end{align}
It is easy to identify the parameters as
\begin{align}
\xi=-i n T \beta\,\qquad \Rightarrow \rho=nT\beta, \quad \theta=-\frac{\pi}{2}\, .    
\end{align}
We can rewrite the state \eqref{eqn:coherent_state_caputa} in the following form
\begin{align}
  e^{-i t L_0}| \Psi(\xi)\rangle&=  e^{-i t L_0} e^{\xi L_{-1}-\bar{\xi}L_1}e^{i t L_0}e^{-i t L_0}|h\rangle \nn
  &=e^{-i  h t}e^{-i t L_0} e^{\xi L_{-1}-\bar{\xi}L_1}e^{i t L_0}|h\rangle\,,
\end{align}
where we have used the notation
\begin{align}
    L_0|h\rangle =h |h\rangle\,.
\end{align}
It is easy to check that
\begin{align}
    e^{a L_0} L_{\pm 1} e^{-a L_0}=e^{\mp a} L_{\pm 1} \,,\quad {\text{for any}}\, \, a\,.
\end{align}

It follows from there
\begin{align}
    e^{-i t L_0} e^{\xi L_{-1}-\bar{\xi}L_1}e^{-i t L_0}= e^{\xi e^{-i t} L_{-1}-\bar{\xi}e^{i t}L_1}\,.
\end{align}
Hence, the role  of $(\xi,\bar{\xi})$ in \eqref{eqn:coherent_state_caputa} is replaced by $(\xi e^{-i t},\bar{\xi} e^{i t})$
\begin{align}
  e^{-i t L_0}| \Psi(\xi)\rangle &=   e^{-i  h t} | \Psi(\xi e^{-i t} )\rangle\,,
\end{align}

which results in the following modified parameter in \eqref{eqn:y}
\begin{align}\label{eqn:coord}
    y=e^{-i \frac{\pi}{2}}e^{-i   t}\tanh{n T \beta}\,.
\end{align}
Plugging \eqref{eqn:coord} in \eqref{eqn:ee_caputa} results in the holographic entanglement entropy ($\sig=n \beta T$) (We ignore $\epsilon_{UV}$ term here).
{\small
\begin{align}
    & S= \frac{1}{6} c \left \lbrace \log \left(-1+e^{-2 i \pi  x}\right)+\log \left(-1+e^{2 i \pi  x}\right)\right \rbrace \\
    & \notag + h\bigg\lbrace 4-2\pi x 
    \cot (\pi  x) +\frac{i}{\tanh^2\sigma-1} \left(\log \left[1+\frac{i \left(\tanh^2 \sig-1\right)}{\tanh \sig  e^{i t}+i}\right]-\log \left[1+\frac{i \left(\tanh \sig^2-1\right)}{\tanh \sig e^{i (t+2 \pi  x)}+i}\right] \right)\\
    & \notag \times \left(2 \tanh \sig\csc (\pi  x) \sin (t+\pi  x)+\left(\tanh^2 \sig+1\right) \cot (\pi  x)\right)\bigg \rbrace\,.
\end{align}
}

The $O(c^0)$ term agrees with the universal contribution $S_{uni}$ in \eqref{eqn:cftuni}.

For computation in the bulk, recall that our boundary interval $A_t$ was at a constant time slice $t$ instead of $t=0$. Hence, the endpoints of the interval in this case are at  $z_1=e^{2\pi a+i t}$ and $z_2=e^{i t+ 2 \pi a+ 2 i \pi x}$. The superposition coefficients in $|\hat{\Psi}\rangle$ are independent of time, hence $y= -i \tanh n T \beta$. With these substitutions in \eqref{eqn:ee_caputa}, the $O(c^0)$ term matches the area term in \eqref{eq: final holo ee}.


\bibliographystyle{JHEP}

\bibliography{ref}

\end{document}